\documentclass[12pt]{JHEP3}

\usepackage{epsfig}
\epsfclipon
\usepackage{multicol}
\usepackage{verbatim}
\usepackage{epsfig,bm}
\usepackage{amssymb,amsmath}
\usepackage{cite}

\newcommand{\roughly}[1]{\mathrel{\raise.3ex\hbox{$#1$\kern-0.85em
\lower1ex\hbox{$\sim$}}}}

\newcommand{\lsim}{\roughly<}
\newcommand{\gsim}{\roughly>}
\newcommand{\sss}{\scriptscriptstyle}

\def\beq{\begin{equation}}
\def\eeq{\end{equation}}
\def\beqa{\begin{eqnarray}}
\def\eeqa{\end{eqnarray}}

\def\cbt{c_{\bar\theta}}
\def\sbt{s_{\bar\theta}}

\makeatletter \@addtoreset{equation}{section} \makeatother

\title{Electroweak Phase Transition and LHC Signatures
in the Singlet Majoron Model}

\author{James M.\ Cline, Guillaume Laporte, Hiroki
Yamashita\footnote{current address: Barclays Capital Japan Limited}\\
McGill University, Department of Physics, Montr\'eal, Qu\'ebec H3A 2T8,
Canada\\
jcline@physics.mcgill.ca, guillaume.laporte@mail.mcgill.ca }

\author{Sabine Kraml \\
Laboratoire de Physique Subatomique et de Cosmologie, UJF Grenoble 1, 
CNRS/IN2P3, INPG, 53 Avenue des Martyrs, F-38026 Grenoble, France\\
sabine.kraml@lpsc.in2p3.fr}

\abstract{
We reconsider the strength of  the
electroweak phase transition in the singlet Majoron extension of the
Standard Model, with a low ($\lsim$ TeV) scale of the singlet VEV.  A
strongly first order phase transition, of interest for electroweak
baryogenesis, is found in sizeable regions of the parameter space,
especially when the cross-coupling $\lambda_{hs}|S|^2|H|^2$ between
the singlet and the doublet Higgs is significant.   Large Majorana
Yukawa couplings of the singlet neutrinos, $y_i S \nu_i^c\nu_i$, are
also important for strengthening the transition.  We incorporate the
LEP and Tevatron constraints on the Higgs masses, and electroweak
precision constraints,  in our search for allowed
parameters; successful examples include singlet masses ranging from 5
GeV to several TeV.  Models with a strong phase transition typically
predict a nonstandard Higgs with mass in the range  $113~{\rm GeV}
\lsim m_H\lsim 200~{\rm GeV}$ and  production cross sections reduced
by mixing with the singlet,  with $\cos^2\theta$ significantly less than 1.   
We also find examples where the singlet is light and the decay $H\to SS$  
can modify the Higgs branching ratios relative to 
Standard Model expectations.}

\preprint{LPSC 09-52}

\begin{document}

\section{Introduction} 

Nonstandard Higgs sectors are interesting from
the perspective of LHC physics and cosmology.  While the Standard
Model predicts a smooth cross-over for the electroweak phase
transition (EWPT) \cite{smooth}, extensions can give a strongly first-order phase
transition, which is a necessary ingredient for electroweak
baryogenesis, and could also possibly generate observably large
gravitational waves \cite{gw} or primordial magnetic fields
\cite{magnetic}.  Supersymmetric extensions of the Standard Model have been
studied the most intensively in this respect \cite{susy},  but it is
also possible to get a strong transition from more generic two-Higgs
doublet models \cite{cl,thdm}, from technicolor theories \cite{tc}, 
higher-dimension operators involving the Standard Model Higgs \cite{hdo}, or from
singlets which mix with the Standard Model Higgs 
\cite{AH,Espinosa:1993bs,Choi:1993cv,Ham:2004cf,Ahriche:2007jp,
Profumo:2007wc,Barger:2008jx,Ashoorioon:2009nf,majoron}.
In the last
category, the singlet Majoron model  \cite{model} is an interesting example since
it was originally motivated by the spontaneously breaking of lepton
symmetry and consequent generation of neutrino masses by the seesaw
mechanism.  It is the model which we consider in the
present work.

In the singlet Majoron  model, right-handed neutrinos $\nu_{{\sss R},i}$ acquire Majorana
masses $M_i = y_i \langle S\rangle$ through their Yukawa couplings
to the complex singlet field $S$, when it gets a VEV.  Denoting the
Yukawa couplings to the doublet Higgs by 
$h_i \bar\nu_{{\sss L},i} H \nu_{{\sss R},i}$,
the seesaw masses of the light neutrinos are given by 
\beq
	m_{\nu,i} = {h^2_i v^2\over y_i \langle S\rangle}
\eeq
where $v$ is the VEV of $H$.  If the Yukawa couplings are  $O(1)$,
then $\langle S\rangle \sim v^2/m_{\nu,i}$, a very high scale 
$\gsim 10^{14}$ GeV.  This is the usual assumption, which would
render the  singlet field irrelevant for physics at the electroweak
scale.  However, we know that small Yukawa couplings exist even in
the Standard Model:  that of the electron is $O(10^{-6})$.  If the
$h_i$ are also of this order (while $y_i\sim 1$), then   $\langle
S\rangle$ could be as small as  $10^{-12} v^2/(0.1{\rm\ eV})\sim 300$
GeV.  From this point of view, a low scale for the singlet is no less
natural than the Standard Model itself, and merits consideration.  

The effect of the singlet Majoron on the EWPT has been  considered
previously in \cite{majoron}, but these papers were written before
the final LEP/Tevatron bounds on the Higgs boson mass $m_H$  or 
values of electroweak precision observables (EWPO) were known, 
and thus they could not
take these important constraints into account.\footnote{It is possible
to evade the EWPO constraints, so we will present
results both with and without applying them.}
  As is well
appreciated, the strength of the EWPT tends to be inversely related
to $m_H$; moreover the EWPO constraints tend to exclude heavy 
singlet fields which have significant mixing with the Higgs doublet.
There have also been studies of related models 
\cite{Espinosa:1993bs,Choi:1993cv,Ham:2004cf,Ahriche:2007jp,
Profumo:2007wc,Barger:2008jx,Ashoorioon:2009nf},  where the
singlet is a real field,  or a complex one such that the global U(1)
symmetry under which $S$ might transform is explicitly broken by
terms like $S^3$.  These models are also very interesting, but
sufficiently different from the Majoron model to justify a separate
study of the latter.  The models with explicitly broken symmetry are
more generic, but not motivated by considerations of neutrino
physics.  Moreover, cubic terms in the scalar potential tend to make
it easier to find a first-order phase transition, so we would expect
the physics leading to a strong EWPT to be qualitatively different in
the two classes of models.  Indeed we will show that the coupling of
the right-handed neutrinos to $S$ plays an important role in getting
a strong phase transition in the  Majoron model.  

Because the global lepton symmetry is spontaneously broken, the
imaginary part of $S$ is a Goldstone boson, the Majoron.  Since we
are assuming the doublet Yukawa couplings to be quite small $(h_i\lsim
O(10^{-6}))$, the massless Majoron couples very weakly to the light
neutrinos, with strength $h_i v/ \langle S\rangle$.  These couplings
are diagonal in the mass basis at this order; off-diagonal couplings
which could lead to neutrino decays are suppressed by  $(h_i v/
\langle S\rangle)^2$ (see ref.\ \cite{couplings} and references
therein).  Such a weakly coupled Majoron
goes out of equilibrium well before nucleosynthesis, and also has a
negligible  effect on energy loss from stars, and so it is
experimentally unconstrained.\footnote{In contrast to the scenarios 
discussed in ref.\ \cite{Barger:2008jx}, we do not require the singlet
to provide a dark matter candidate.}  

In the remainder of the paper, we derive the finite-temperature
effective potential (section \ref{potsec}), and present our methods
and results  for the strength of the EWPT from a wide search of the
model's parameter space (section \ref{results}).  We analyze the
nonstandard decay modes and 
discovery potential of the singlet sector at the LHC in section
\ref{LHCsect}.  Conclusions are
given in section \ref{conclusions}.  We present formulas for
field-dependent and thermal masses needed for the potential in the
appendices, as well as those pertaining to our renormalization
prescription, the running of the couplings, and formulas for the
oblique parameters for electroweak precision observables.

\section{The potential}

\label{potsec}
Because the left-handed neutrino Yukawa couplings are assumed to be
very small for our purposes, we neglect them in what follows. 
Similarly, only the top quark is retained amongst the other fermions
of the Standard Model.  At tree level the  potential is then
\beq
	V_0 = {\lambda_h}\left( |H|^2- \frac12v_h^2\right )^2+
	{\lambda_s}\left( |S|^2- \frac12v_s^2\right )^2
	+\lambda_{hs} |S|^2|H|^2  
	+ y_t\bar Q Ht_r + \frac12\sum_i y_i S\nu_i\nu_i + {\rm h.c.}
\label{V0}
\eeq
in terms of the complex Higgs doublet $H = (H^0,H^+)$,
complex singlet $S$, top quark and right-handed neutrinos.  
For definiteness, we take three generations of right-handed neutrinos
with equal Majorana Yukawa couplings  $y_i$.
Due to the
cross-coupling $\lambda_{hs}$, $v_h$ and $v_s$ are not generally the VEV's
of the fields at the minimum of the potential.  Rather, the relation
is
\beqa
	v_s^2 &=& 2\langle S\rangle^2 + {\lambda_{hs}\over\lambda_s}
	\langle H\rangle^2; \nonumber\\
	v_h^2 &=& 2\langle H\rangle^2 + {\lambda_{hs}\over\lambda_h}
	\langle S\rangle^2;
\label{vrels}
\eeqa
We take $\langle H\rangle\cong 174$ GeV and $\langle S\rangle$ (to be
varied) as the
physical input parameters.  Because of the $Z_2$ symmetries $H\to -H$
and $S\to -S$ in the scalar potential, there is no loss in generality in assuming that both
VEV's are positive. (The signs of the fermion masses are not
physically significant.)

At finite temperature, the lowest order thermal correction to $V_0$
is a function of the field-dependent particle masses, $m_i(H,S)$:
\beq
	\Delta V_{T} = T \sum_i\pm \int {d^{\,3}p\over (2\pi)^3}
	\ln\left(1 \mp e^{-\beta\sqrt{p^2+m^2_i(H,S)}}\right)
	\ \left\{ {\hbox{bosons}\atop\hbox{fermions}}\right.
\label{v1l}
\eeq
where $\beta = 1/T$.  
These functions are often approximated by their high temperature
expansions, but for numerical purposes it is preferable to use an
approximation that works at all values of $m_i/T$.  We use the
approximation described in ref.\ \cite{cl}, in which the high-$T$
and low-$T$ expansions are smoothly joined together at some large
value of $m_i/T$.  This is reviewed in appendix \ref{appA}.  The
expressions for the field-dependent thermal masses are given in 
appendix \ref{appB}.

Furthermore, it is important to improve the thermal contribution by
resumming the ring diagrams, which amounts to replacing $m_i^2$ with
the thermally corrected masses (of the form $m_i^2 + c_i T^2$) in
eqs.\ (\ref{v1l}).  Otherwise there is a danger of overestimating the
strength of a first order phase transition.   Often, this
substitution is made only in the cubic term of the high-$T$
expansion, where it has the biggest effect on the barrier between the
true and false vacua of the potential \cite{AE}.  However, when one uses an
expression that correctly captures both the large and the small
$m_i/T$ behavior of (\ref{v1l}), there is no way to consistently
include thermal mass effects only in the cubic term, since it appears
explicitly only in the high-$T$ expansion.   Trying to do so creates a kink in 
the potential when the high-$T$ and low-$T$ expansions
are joined onto each other. 
 To avoid such complications, we simply replace $m_i^2$
by its thermally corrected value everywhere in (\ref{v1l})
\cite{Parwani}.   The discrepancy between the two approaches is
formally significant only 
when computing the potential to two loops.

The thermal potential (\ref{v1l}) can be regarded as a one-loop effect, so
for consistency one must also include the one-loop, zero-temperature
(Coleman-Weinberg) correction to the potential, 
\beq
\Delta V_{CW} = \frac12 A|H|^2 + {1\over 64\pi^2}\sum_i m_i^4(H,S)\left(\ln{m_i^2(H,S)\over
\mu^2} -\frac32\right)\times
	\ \left\{ {\hbox{$+1$, bosons$\phantom{a}$}\atop\hbox{$-1$, fermions}}\right.
\label{CW}
\eeq
where $A|H|^2$ is a counterterm and $\mu$ is the renormalization
scale.  We do not include a counterterm of the form $B|S|^2$ because
this can be absorbed into a redefinition of $v_s$.  However for the
doublet Higgs it is convenient to introduce the $A|H|^2$ counterterm, because
then one can maintain the tree-level  relation between the Higgs VEV
$\langle H\rangle$ and $v_h$.  In fact, it is convenient to maintain
both relations (\ref{vrels}), so that the position of the
zero-temperature minimum of the potential $H=\langle H\rangle$,
$S=\langle S\rangle$, is known analytically.  We thus adopt as our
renormalization prescription
\beq
	{\partial V\over\partial H} = {\partial V\over\partial S} = 0
\ \hbox{\ at \ \ $H=\langle H\rangle$, $S=\langle S\rangle$}
\label{renconds}
\eeq
where $V = V_0 + \Delta V_{CW}$.
These two equations can be analytically solved to find $\mu$ and $A$
(see appendix \ref{appC}).  

A notable feature of $\Delta V_{CW}$ is that the terms
$\ln(m^2_i(H,S))$ appear in such as way as to exactly cancel
corresponding terms in the high-$T$ expansion of $\Delta V_{T}$; then
the cubic term $(m^2_i)^{3/2}$ of $\Delta V_{T}$ is the only source
of nonanalytic dependence on the fields.  To preserve this property,
we also replace $m^2_i$ by the thermally corrected expression in
$\Delta V_{CW}$ when we do the ring improvement of the
potential \cite{Parwani}.

A complication which arises in the effective potential is the
appearance of negative values of $m^2_i(H,S)$ for the Goldstone boson
degrees of freedom at small $H$ or $S$; this can happen even when the
thermal correction to $m^2$ is included.  Such values create a problem
with the cubic term $(m^2_i)^{3/2}$ in the high-$T$ expansion of the
thermal potential.  Even if one
takes only the real part, which vanishes for negative $m^2$,
derivatives of this with respect to the fields are discontinuous at
the point where $m^2$ changes sign, leading to serious difficulties
for algorithms which attempt to minimize the potential.
There are various prescriptions in the literature for dealing with
the Goldstone bosons.  We take the simplest approach, which is to
simply omit their contributions from $\Delta V_{CW}$ and 
$\Delta V_T$.  Experience with other models indicates that the
Goldstone bosons never have a strong effect on the phase transition
in any case.  Thus we omit the contributions from $H^\pm$, Im($H^0$), and
the Majoron $j$ in the sums.

In the limit of very heavy singlet and neutrinos, one expects the 
effects of the new physics to decouple.  This is evident in the
low-$T$ expansion of the thermal correction to the potential, since
the effects of heavy particles are Boltzmann suppressed.  However,
this decoupling is missing from the naive thermal corrections to the masses,
(\ref{thermass1}-\ref{thermass1}), which were derived from the high-$T$ expansion of
the potential.  To correct for this, we insert Boltzmann 
factors involving the heavy particle masses,
\beqa
	\delta m^2_{H} &=& T^2\left(\frac12\lambda_h +
\frac1{12}\lambda_{hs}
e^{-m_S/T}
	+ \frac1{16}(3g^2 + g'^2) + \frac14 y_t^2\right) \\
\label{thermass1a}
	\delta m^2_{S} &=& T^2\left(\frac13\lambda_s +
\frac1{6}\lambda_{hs}
	+ \frac{1}{24}\sum_i y_i^2 e^{-m_\nu/T}\right)
\label{thermass2a}
\eeqa
where $m_S$ and $m_\nu$ are evaluated at zero temperature and at
the minimum of the potential.  This procedure is somewhat rough,
but better than ignoring the issue altogether \cite{CE}.

\section{Search of parameter space}\label{results}
Our goal was to make a broad scan of the parameter space, in search of
models giving a sufficiently strong EWPT for electroweak baryogenesis.
We used a $30\times 15^2\times 20\times 30$ grid ($\sim 4\times 10^6$ points) 
on the five parameters
$\lambda_{hs}$, $\lambda_h$, $\lambda_s$, $y_i$ (taking equal Majorana
Yukawa couplings for 3 generations of right-handed neutrinos) and
$\langle s\rangle$, the VEV of the real component of $S$, 
$\langle s\rangle = \sqrt{2}\langle S\rangle$.  After a preliminary scan of the parameter
space to determine the values of interest, 
these were taken to be in the ranges
\beq
	0 < \lambda_s,\lambda_h < 3,\quad
	-3 < \lambda_{hs} < 3,\quad
	0 < y_i^2 < 8,\quad 
	\langle s\rangle < 1800  {\rm\ GeV}
\label{ranges}
\eeq
subject to the constraint $\lambda_{hs} > -\sqrt{\lambda_h\lambda_s}$ which
is needed for $V$ to be bounded from below for large field values (at tree
level).
The chosen ranges include large values of the coupling constants, but since
$\lambda_i^2/4\pi, y_i^2/4\pi \lsim 1$, they are not unreasonably large.

\FIGURE[ht]{
\epsfig{file=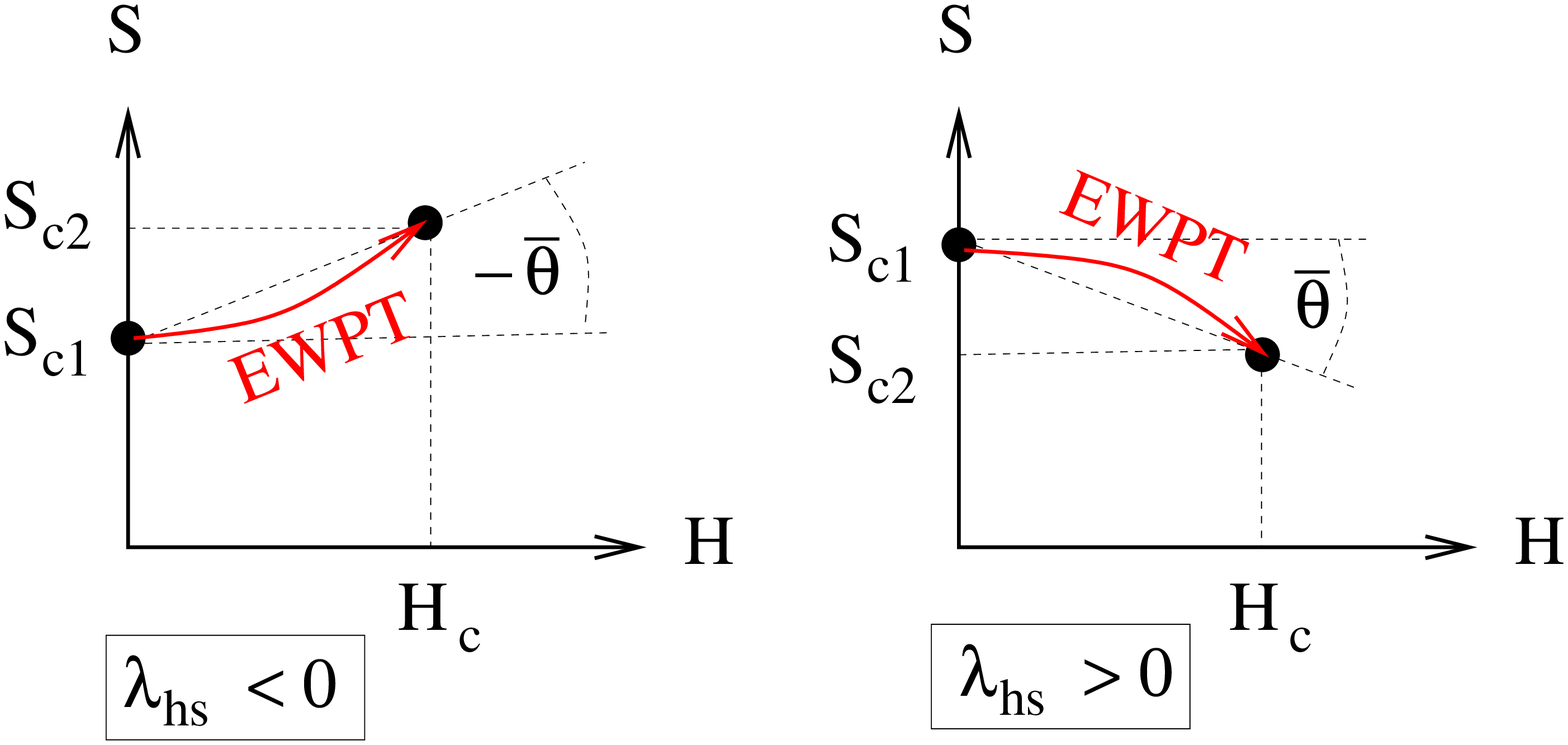,width=0.7\hsize}
\caption{Trajectories of typical EWPT in $H$-$S$ field space.
The significance of the mixing angle $\bar\theta$ is discussed near eq.\ (3.9).}
\label{trans}
}

We find that the typical pattern of symmetry breaking is to
first develop a finite-temperature VEV at $S=S_{c1}$ for $S$ alone as $T$ is
lowered from very high values, {\it i.e.,} the phase transition for 
the singlet to condense usually occurs above the EWPT.  
The electroweak
transition is a jump from this false vacuum along the $S$ axis to a
true vacuum in which $H\neq 0$;  furthermore $\langle S\rangle=S_{c2}$
increases relative to its value in the  $\langle H\rangle=0$
minimum if $\lambda_{hs}<0$, and decreases if $\lambda_{hs}>0$.
This pattern is shown in figure \ref{trans}.  Although
$S_{c1}$ is typically large, we will see that it can sometimes
(when $\lambda_{hs}<0$) be zero.

 When $|\lambda_{hs}|< 1$, the mixing between the $H$ and
$S$ fields is small, and the transition takes place mostly in the $H$
direction.  For larger values of $|\lambda_{hs}|$, which are more
typical of cases with a strong EWPT, the induced $H$-$S$ mixing
causes the tunneling path to be along a linear combination of the
fields.  This is illustrated in figure  \ref{potfig}.  
It is easy to understand the shapes of the valleys in the potential
indicated in figs. \ref{trans},\ref{potfig}.  First, the discrete symmetry 
$H\to-H$ implies that the light direction will be purely parallel
to the $H$ axis at $H=0$.  When $H$ gets a VEV, the term
 $\lambda_{hs}|H|^2|S|^2$ makes the squared mass of $S$ more negative
if $\lambda_{hs}<0$, increasing the VEV of $S$ in the electroweak
symmetry breaking vacuum.  Conversely the VEV of $S$ is decreased for
$\lambda_{hs}>0$.

The above discussion makes it possible to understand why the mixing
term $\lambda_{hs}|H|^2|S|^2$ can generally strengthen the EWPT. 
Roughly, we expect the VEV of $H$ to scale like the square root of
$-\mu_h^2$, the negative mass squared term for $H$.  
When $S$ has a VEV, $-\mu_h^2$ gets a contribution 
$\lambda_{hs}\langle S\rangle^2$.  Now suppose that 
$\langle S\rangle^2$ changes by the amount 
$\delta\langle S\rangle^2$ when $H$ makes the transition between the
symmetric and electroweak symmetry breaking EWSB  vacua.  We expect that the critical value $H_c$
increases with $\lambda_{hs}\,\delta\langle S\rangle^2$.  The preceding
discussion shows that this quantity is always positive, regardless of
the sign of $\lambda_{hs}$, so the mixing term should tend to
strengthen the EWPT whenever it has a significant size.

\FIGURE[t]{
\centerline{\epsfxsize=\textwidth\epsfbox{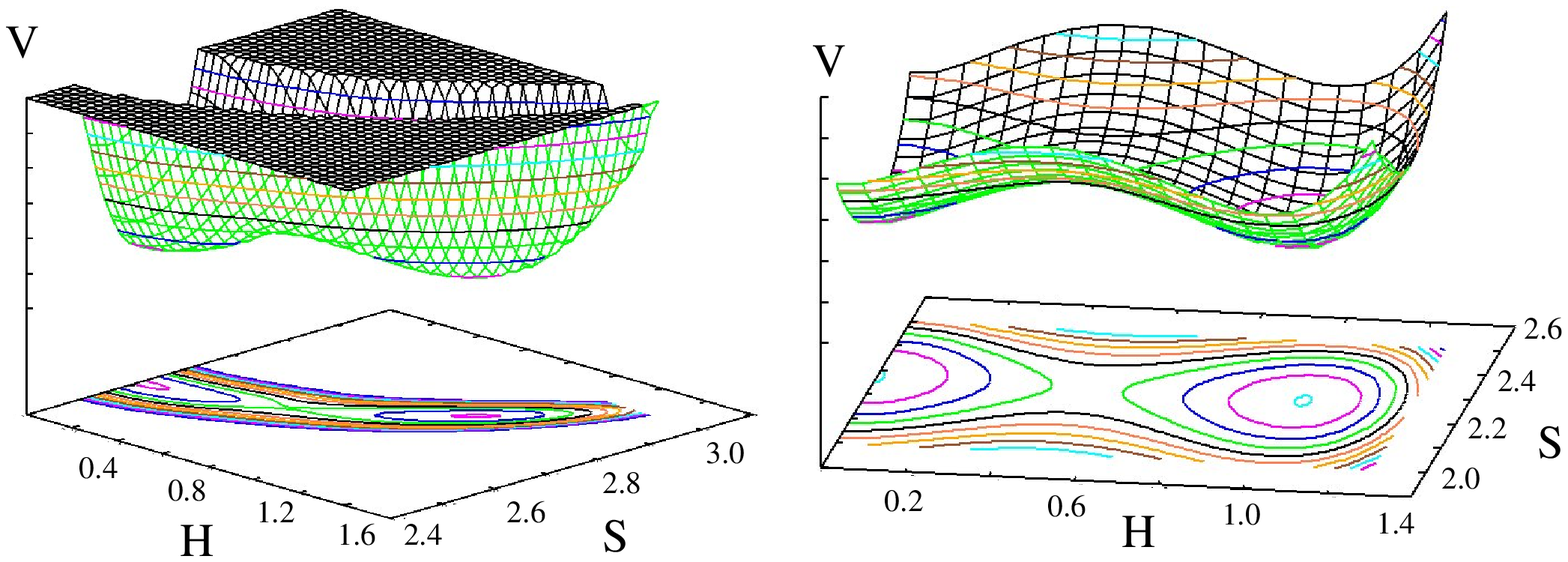}}
\caption{Shape of potential $V(H,S)$ at the critical temperature for
large value of $\lambda_{hs}=-2.4$ (left) and small value
$\lambda_{hs}=-0.2$ (right), illustrating the effect of $\lambda_{hs}$
on the path in field space between the degenerate minima.  Contours of
the potential are projected onto the lower plane.  $H$, $S$ are in
units of 100 GeV.
}
\label{potfig}
}

\subsection{Algorithm} Because the phase transition typically
proceeds in two steps, due  to the two condensing fields, automating
the search for a strong first order transition proved to be somewhat
more difficult in this model than for effectively single-field
models.  The key steps are to identify whether there is a
barrier between the trivial $\langle H\rangle =0$ minimum and the
EWSB $\langle H\rangle \neq 0$ minimum, and to bracket the critical
temperature if there is one.  Figure \ref{flowchart} outlines our
algorithm.  It outputs a logical variable {\tt success} to indicate
whether a first order transition with $v_c/T_c > 1$ was found, and
bracketing temperatures {\tt Tmin} and {\tt Tmax} if so.

\FIGURE[t]{
\centerline{\epsfxsize=0.75\textwidth\epsfbox{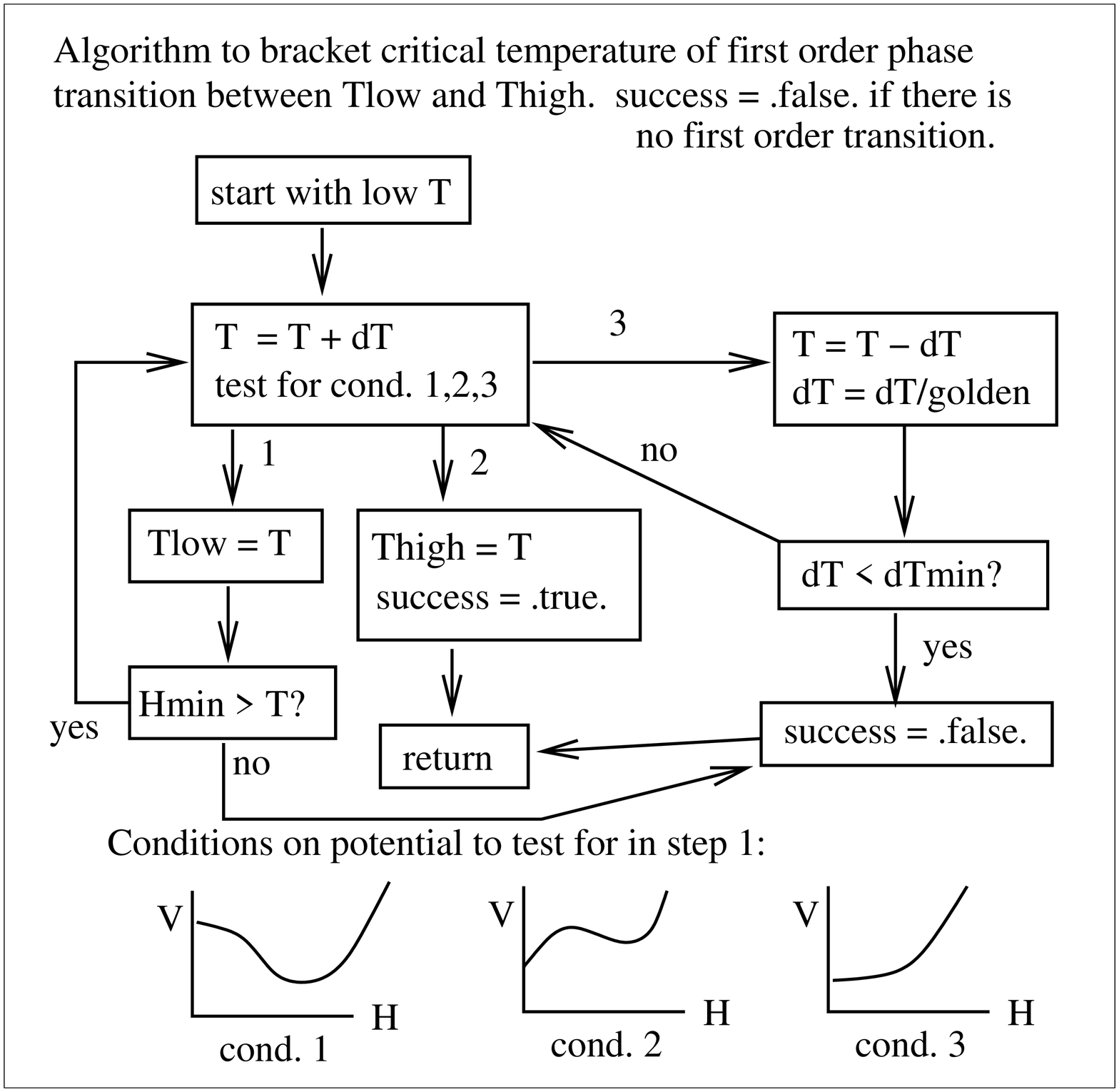}}
\caption{Flowchart for algorithm to bracket the critical temperature
for a first order phase transition.}
\label{flowchart}
}

The algorithm shown in figure \ref{flowchart} is
simple to implement in a single-field model, but when there are two
fields it can be difficult to properly identify the relevant
direction along which to check the curvature of the potential.  In
the present case, we start by finding the global electroweak
symmetry  breaking minimum near $T=0$, and a local $S$-breaking
minimum or  saddle point on the $S$ axis, and raise the temperature
until these two critical points become degenerate.  Naively, the
tunneling path would be along a line connecting these two points,
but because of the ``banana'' effect shown in fig.\ \ref{potfig}, the
relevant flat direction might really be curved.  One must therefore
determine the curvature  locally around the two putative minima to
check that they really are minima.  If the transition is actually
second order but the transition path is curved, one could mistakenly
conclude that it is first order by measuring the curvature of the
potential along the straight line rather than the curved path.  Of
course a visual inspection of the shape of the potential would
eliminate such cases, but we needed to automate this.  To do so, we
minimize the potential on small circles surrounding the putative
local minima, to verify that they indeed are minima, and to find the
directions of shallowest ascent.

\subsection{Criteria for accepted points}

The basic requirement for a strong enough phase transition is
\beq
	{v_c\over T_c} \gsim 1
\eeq
where $v_c$ is the VEV of the real Higgs field $h$ ($\sqrt{2}\times
174\cong 246$ GeV
at zero temperature) at the critical temperature $T_c$ \cite{Shap}
(see \cite{JC} for a pedagogical review).  This avoids the washout of baryons
produced during the EWPT by sphalerons.   A more careful treatment
would be to calculate the sphaleron energy in the model at hand, since
this can in principle be different from the Standard Model value and change the 
bound.  The change is typically small however, and so we do not
consider this effect.  

\FIGURE[t]{
\centerline{\epsfxsize=0.55\textwidth\epsfbox{lep.eps}}
\caption{95\% c.l.\ LEP bound on mixing angle $(\cos^2\theta$) of the 
mostly-doublet Higgs state, from table 14 of ref.\ \cite{LEP}.  
The same bound applies to 
$\sin^2\theta$ for the mostly-singlet state.}
\label{lepfig}
}

In addition, we demand that the LEP limit on the Higgs mass \cite{LEP}
be satisfied.  In this regard, another important feature of the
$\lambda_{hs}|H^2||S^2|$ interaction is that it can cause large mixing
between the singlet and doublet Higgs bosons, leading to a reduction
in the production cross section for the lightest mass eigenstate
\cite{pheno}.  
We take the fluctuations of the flavor and mass eigenstates to be 
related via
\beq
	\left(\begin{array}{c} \delta H\\ \delta S\end{array}\right) = 
	\left(\begin{array}{cc} $\phantom{-}$\cos\theta & \sin\theta \\
			-\sin\theta & \cos\theta \end{array}
	\right) \left(\begin{array}{c} \delta H'\\ \delta S'\end{array}\right)
\label{mixing}
\eeq
We restrict $\cos\theta\ge 1/\sqrt{2}$, so that $H'$ is the
``Higgs-like'' state and $S'$ is the ``singlet-like'' state,
regardless of which one is heavier.  The mixing angle suppresses the 
couplings of either state
relative to the couplings of a SM Higgs boson. 
The production cross section of the Higgs-like state is reduced by 
$\cos^2\theta$, while that of the singlet-like state scales like
$\sin^2\theta$.  We demand that both of these are less than the LEP
limit;  {\it i.e.,} both $\cos^2\theta$ and $\sin^2\theta$, evaluated
at the appropriate mass, must be less than the value in column (a) of table 14
of ref.\ \cite{LEP}.  The bound is shown in figure \ref{lepfig}.  
The procedure of applying the same bound independently
to both states must be modified if the two are close to 
each other in mass.  However we will find that for cases that give
rise to a strong EWPT, there is always a large separation between the
masses, justifying this simpler approach.\footnote{In particular,
we find no cases where both $m_{H'}$ and $m_{S'}< 114.4$ GeV; in this
situation the decays $H'\to S'S'$ or $S'\to H'H'$ (if kinematically
allowed) could modify
the branching ratios with respect to the Standard Model prediction.}

Recently, the D0 and CDF experiments have disclosed new limits
excluding the SM Higgs boson in the region $160-170$ GeV
\cite{tevatron}.  We fit the limit on $\cos^2(\theta)$ from 
Table XIX in this region with the quadratic function 
$
	\cos^2\theta_{\rm max} = 142.43 - 1.716\, m_{H'} + 
	0.0052\, m_{H'}^2
$
where $m_{H'}$ is in GeV.  This has values $0.99,\,0.86,\,0.99$
at $m_{H'}=160,\,165,\,170$.  Because of this limited range, the
effect is small in our first search of parameter space, which
involves a large range of couplings.  However in the search which
targets smaller couplings, the CDF/D0 limit covers a larger fraction
of the range of allowed masses, and the mixing angles tend to
be smaller, and so the constraint has a more pronounced effect:
450 out of 5300 parameter sets are removed. 

Some cases of interest have very light singlets.   Ref.\
\cite{lightS} noted that values of $m_{S'} \lsim 5$ GeV  (the $B$
meson mass)  are strongly constrained ($\theta<10^{-2}$)  by the
decays $B\to S' X$ followed by $S'\to \mu^+\mu^-$.  We thus exclude
$m_{S'} \lsim 5$ GeV if the mixing angle is greater than $0.01$.
This has a negligible effect on our broader search of the parameter
space since a very small fraction of this sample has light singlets,
but in the search which is limited to smaller values of the
coupling constants, this constraint is more significant.

Another important criterion is that the EWSB vacuum at $T=0$ must exist.  
Although this seems obvious, our
broad  scan of parameter space includes cases where, due to
radiative corrections to the tree-level Higgs potential, the
curvature of the potential is positive when $H=0$, leading to no
EWSB.  In fact, we will see that some concentrations of the preferred
parameter space tend to be close to this perilous edge, especially
when $\lambda_{hs} <0$.

Since we consider models with large couplings and large
masses, a consistency requirement for perturbation theory to be under
control is that none of the running couplings diverge (reaching a
Landau pole) at renormalization scales smaller than the heaviest
particle masses.  The beta functions for the largest couplings
are given in appendix \ref{betafns}.  For each otherwise accepted
parameter set, we integrate these to find the first Landau pole
and discard parameters which fail this test.  This eliminates 
approximately 15\% of otherwise accepted parameter sets from the
range (\ref{ranges}).

Finally, we include constraints on the oblique parameters $S$, $T$,
$U$ from electroweak precision observables (EWPO).  In order to not
to mask the intrinsic dynamics of the phase transition too much, we
chose to first present results without inclusion of the EWPO
constraint.   A separate section \ref{ewpo} is devoted to showing how
the results are affected by its inclusion.  We give details about its
implementation there and in appendix \ref{EWPOS}.

\subsection{Distributions of parameters}

\FIGURE[b]{
\centerline{\epsfxsize=\textwidth\epsfbox{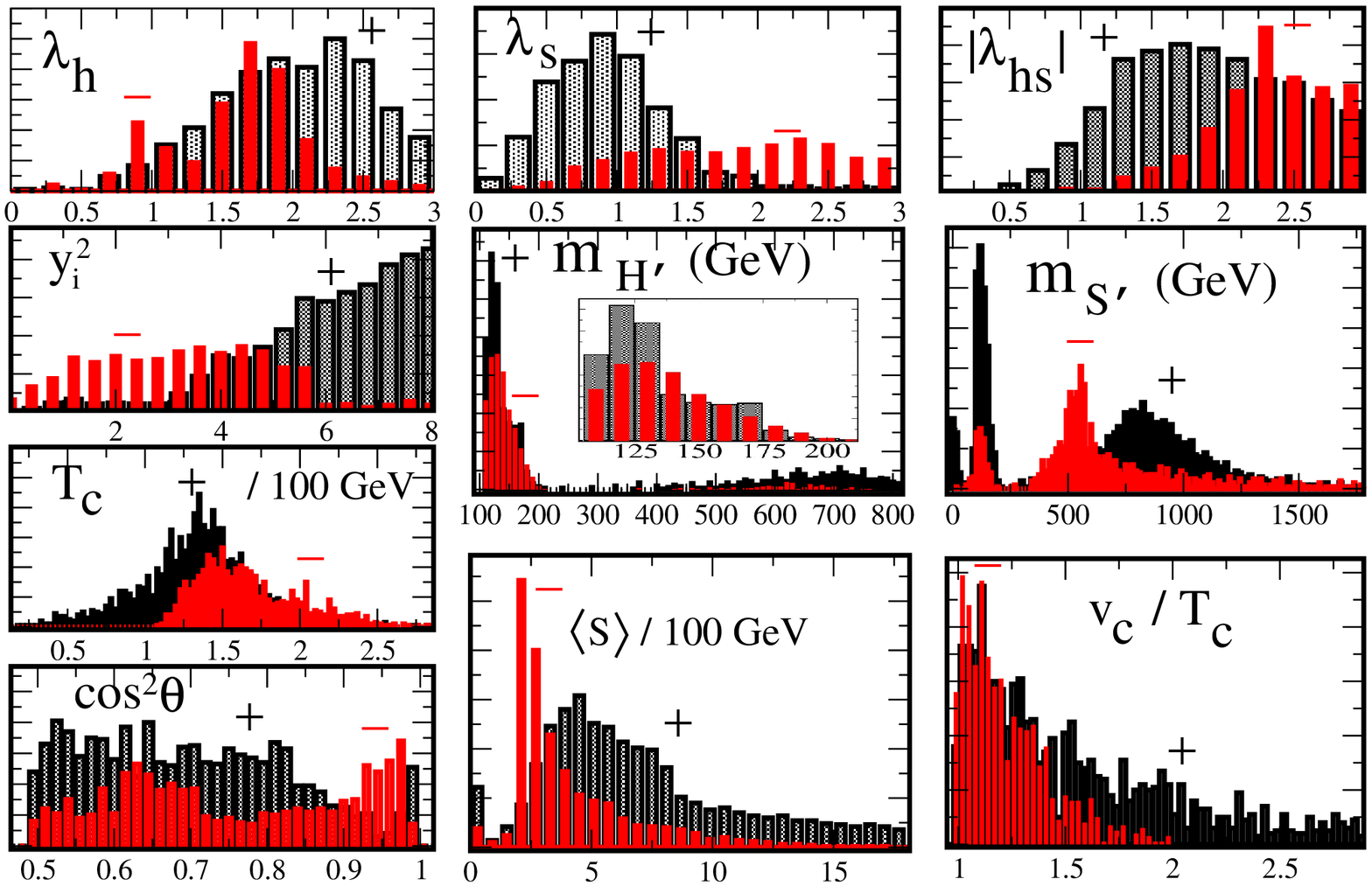}}
\caption{Distributions of parameters which generate a strong
enough first order phase transition.  Lighter (red) bars correspond
to $\lambda_{hs}<0$, darker (black) to $\lambda_{hs}>0$, and
$|\lambda_{hs}|$ is shown in the distribution for $\lambda_{hs}$.
$\pm$ sign indicates regions associated with
$\lambda_{hs}\,{\scriptstyle\gtrless}\,0$.  $\langle s\rangle$ is
the real (not complex) singlet VEV.
}
\label{hists}
}

Out of the $4\times 10^6$ points tested on our uniform grid in the
parameter space, approximately $0.07\%
$ generate a strong enough phase transition and fulfill the other
criteria mentioned above.  We display the
distributions of accepted parameters in figure \ref{hists}.  The
samples are divided into two groups, according to whether
$\lambda_{hs}<0$ or $\lambda_{hs}>0$, due to the expected qualitative
differences between the two cases.  These differences are highlighted
by the separate distributions shown for $\lambda_{hs}<0$ and 
$\lambda_{hs}>0$.  One feature which they have in common however is
the need for generally large values of $|\lambda_{hs}|$, in agreement
with our argument that $H$-$S$ mixing is important for boosting the
strength of the phase transition.

Another striking feature of the distributions is the preference for
large values of the Majorana neutrino Yukawa coupling, $y_i^2$, and
the largest values being correlated with $\lambda_{hs}>0$.  It was
pointed out in ref.\ \cite{Carena} that new heavy fermions 
with a large Yukawa coupling to the Higgs could strengthen the EWPT.
One might wonder whether this could be the origin of the need for
large $y_i^2$ in our model, since both VEV's $S$ and $H$ are changing
during the phase transition.  However, the sign goes the wrong way.
The heavy fermion effect requires the fermion to be heavier in the
EWSB phase than in the symmetric phase.  Fig.\ \ref{trans} shows that
this is the case when $\lambda_{hs}<0$, but not for $\lambda_{hs}>0$.
However the preference for large $y_i^2$ is greater for $\lambda_{hs}>0$
in fig.\ \ref{hists}.

To understand this behavior, we have varied $y_i^2$ away from the
accepted value for some sample points.  We typically find that
$v_c/T_c$ is an increasing function of $y_i^2$, while the light Higgs
mass is decreasing.  There is thus a tension between the demand for a
strong phase transition and the LEP bound, which results in a narrow
window of allowed $y_i^2$ values,  while keeping the other couplings
fixed.  These dependencies on $y_i^2$ are illustrated around a sample
accepted point in fig.\ \ref{yi2ls}(a).

\FIGURE[t]{
\centerline{\epsfxsize=0.5\textwidth\epsfbox{ydep.eps}
\epsfxsize=0.5\textwidth\epsfbox{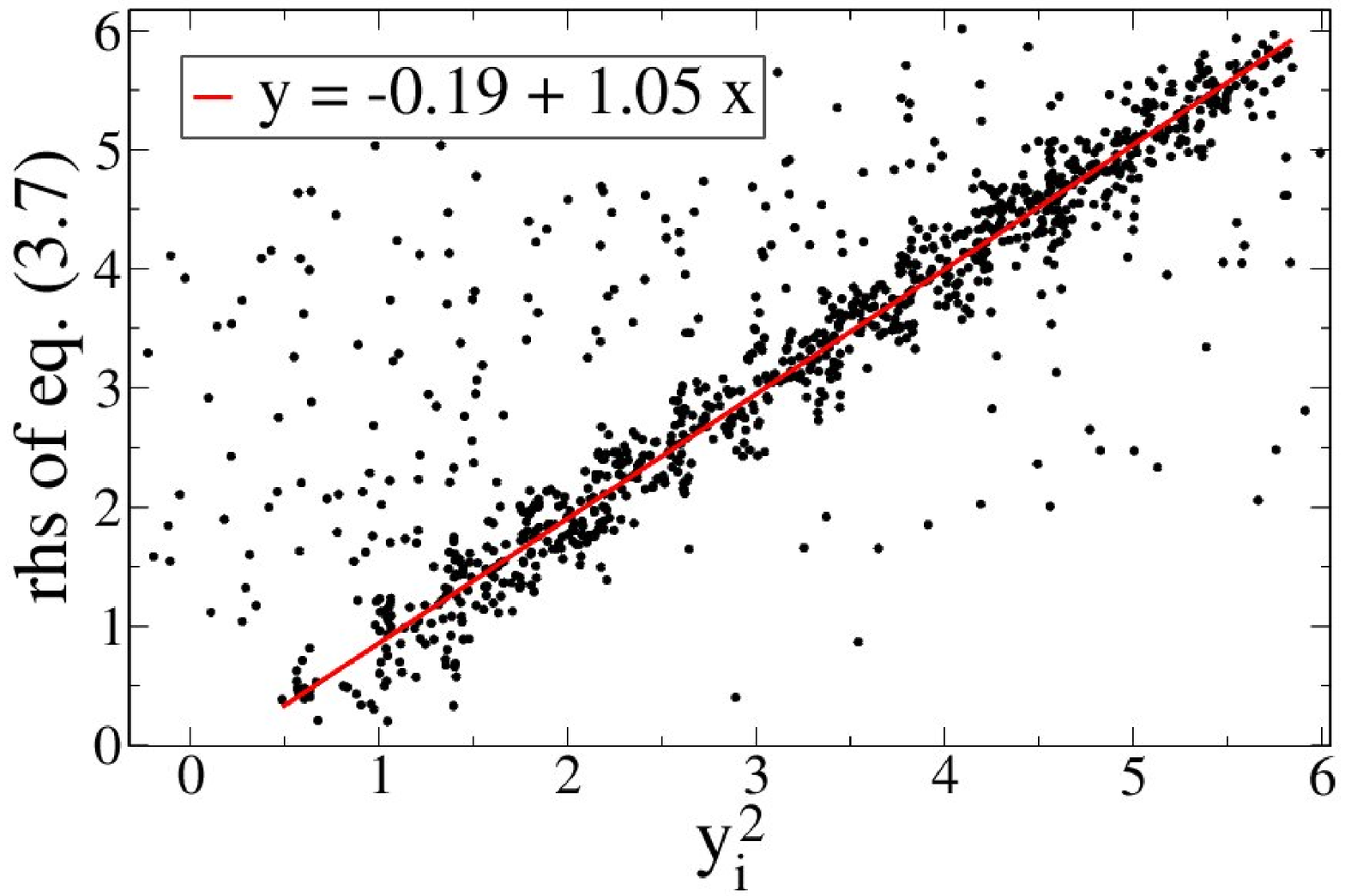}}
\caption{(a) Left: $T_c$, lightest Higgs mass $m_{``H"}$ and $v_c/T_c$ as a
function of $y_i^2$, for representative point 
$\lambda_{hs} = 1.90$, $\lambda_h =1.70$, $\lambda_s = 0.90$,
$\langle s\rangle = 570$ GeV.  (b) Right: the r.h.s.\ of eq.\
(3.7) versus
$y_i^2$ for parameter sets leading to a strong EWPT.
Solid line is the fit to the cluster of points which nearly satisfy
this linear relation.
}
\label{yi2ls}
}

By testing many hypotheses, we eventually discovered an analytic explanation
for the trends visible in the distributions of parameters in fig.\ 
\ref{hists}.  It depends crucially on the one-loop zero-temperature
correction to the effective potential. Let us try to make a rough analytic
estimate of the strength of the phase transition, which is
characterized by $v_c/T_c$.  The critical temperature is approximately
where the temperature-dependent mass squared of $H$, evaluated at
$(H,S)=(0,S_{c1})$, goes through zero.   Using the
field-dependent mass $m^2_{hh}$ of (\ref{masses}), the one-loop
counterterm $A$ of (\ref{CW}), and the  temperature correction
(\ref{thermass1}), we get $T_c^2\sim ({m^2_{hh}(0,S_{c1})-A)/(\delta
m^2_{hh}/T^2)}$.  Putting these results together gives the estimate
\beq
	{T_c}\ \sim\ {
	\sqrt{\lambda_h\langle H\rangle^2 - \lambda_{hs}(S_{c1}^2 -
	\langle S\rangle^2) - A}
\over \sqrt{\frac12\lambda_h +\frac1{12}\lambda_{hs} +\frac14}
}
\label{Tc}
\eeq
The critical temperature can become small relative to $v_c$ if the
Higgs mass renormalization constant $A$, eq.\ (\ref{bigA}), becomes 
large.  This can happen when the renormalization scale
$\ln\mu^2$, eq.\ (\ref{lnmu2}), becomes large.  It is straightforward
to show that  
\beq
	\ln\mu^2 \sim {O(\lambda_s^2,y_i^4,\lambda_{hs}^2)\,
	\langle S\rangle^2 
	+\lambda_{hs} O(\lambda_{hs},\lambda_h,\lambda_{s})\,
        \langle H\rangle^2\over
(24\lambda_s^2-6 y_i^4 +\lambda_{hs}^2)\,\langle S\rangle^2
	+\left(\lambda_{hs}^2+\lambda_{hs}(4\lambda_h+6\lambda_s)
	\right)\langle H\rangle^2}
\label{lnmutoo}
\eeq
where the numerator is correct in order of magnitude,
 while the denominator is
exact.  Since $\ln\mu^2$ appears in $A$, eq.\ (\ref{bigA}), large values of
$\ln\mu^2$ can cause $A$ to be large,
\beq
	A \sim {\ln\mu^2\over
16\pi^2}\left(O(\lambda_h^2,+\lambda_{hs}^2)\langle H\rangle^2
	+
	\lambda_{hs}O(\lambda_s,\lambda_h,\lambda_{hs})\langle S\rangle^2\right)
\eeq
The denominator of eq.\ (\ref{lnmutoo}) vanishes when the relation
\beq
	y_i^2 = \left(4\lambda_s^2 -\frac16\lambda_{hs}^2 +
{\langle H\rangle^2\over 6 \langle S\rangle^2}(\lambda_{hs}^2 +
	4\lambda_h\lambda_{hs})\right)^{1/2}
\label{linear-corr}
\eeq
is satisfied.
The correlation between $y_i^2$ and the r.h.s.\ of  eq.\
(\ref{linear-corr}) is shown in fig.\ \ref{yi2ls}(b). This
relation is just what one would expect from trying to minimize $T_c$
by making $\ln\mu^2$, hence $A$, large.  For smaller $y_i^2$, 
the strength of the
transition rapidly diminishes.  For larger values, we lose the EWSB
vacuum because the curvature $m^2_{hh}$ has the wrong sign at $H=0$.

\FIGURE[t]{
\epsfig{file=tscdist-n.eps,width=0.5\hsize}
\caption{Distribution of $S_{c1}$, the scalar VEV in the electroweak
symmetric vacuum at the critical temperature.}
\label{tscdist}
}

This effect also allows us to understand other features of the
distributions shown in fig.\ \ref{hists}.   Eq.\
(\ref{linear-corr}) shows that when $\lambda_{hs}>0$, larger values
of $y_i^2$ result due to the term $4\lambda_h\lambda_{hs}$; this
trend is seen in the histogram for $y_i^2$.  Similarly, this term
puts a limit on the  magnitude of $\lambda_h$ when $\lambda_{hs}<0$
but not when  $\lambda_{hs}>0$, in agreement with the histogram for
$\lambda_h$. Moreover, inverting the relation (\ref{linear-corr}) to
express $\lambda_s$ in terms of the other variables readily explains
why larger values of $\lambda_s$ are favored for $\lambda_{hs}<0$.
Solving for $\langle H\rangle^2/\langle S\rangle^2$ similarly shows
why larger values of $\langle S\rangle$ occur for $\lambda_{hs}>0$.
In short, the relation (\ref{linear-corr}) allows us to qualitatively
understand most of the trends exhibited in fig.\ \ref{hists}.  The
Coleman-Weinberg potential thus plays an important role in
strengthening the EWPT.  We have further tested this conclusion by
running our program with the one-loop zero-temperature correction 
turned off, finding that the number of accepted points is drastically
reduced in this case. 

We mentioned earlier that the VEV of $S$ in the electroweak
symmetric vacuum, $S_{c1}$, can also be zero, but only when
$\lambda_{hs} < 0$.  The distribution of $S_{c1}$ 
in figure \ref{tscdist} indeed shows a spike 
at $S_{c1}=0$ for $\lambda_{hs} < 0$.  Fig.\ \ref{trans}
makes it clear why there such is a correlation between $S_{c1}$ 
and $\lambda_{hs}$: for $\lambda_{hs} > 0$, $S$ must decrease during
the transition.  It cannot do so if it is already zero.  Eq.\
(\ref{Tc}) also gives insight into this correlation: large values
of $S_{c1}$ are disfavored when $\lambda_{hs}<0$ since this tends to
increase $T_c$ and decrease $v_c/T_c$.

\subsection{Regime of smaller couplings}

\FIGURE[b]{
\centerline{
\epsfxsize=0.47\textwidth\epsfbox{rge.eps}
\epsfxsize=0.5\textwidth\epsfbox{nrge.eps}}
\caption{Distributions of maximum renormalization scale, where
Landau pole develops, for the large-coupling parameter sets (left
panel) and the smaller-coupling ones (right panel).}
\label{rge}
}

\FIGURE[t]{
\centerline{
\epsfxsize=0.5\textwidth\epsfbox{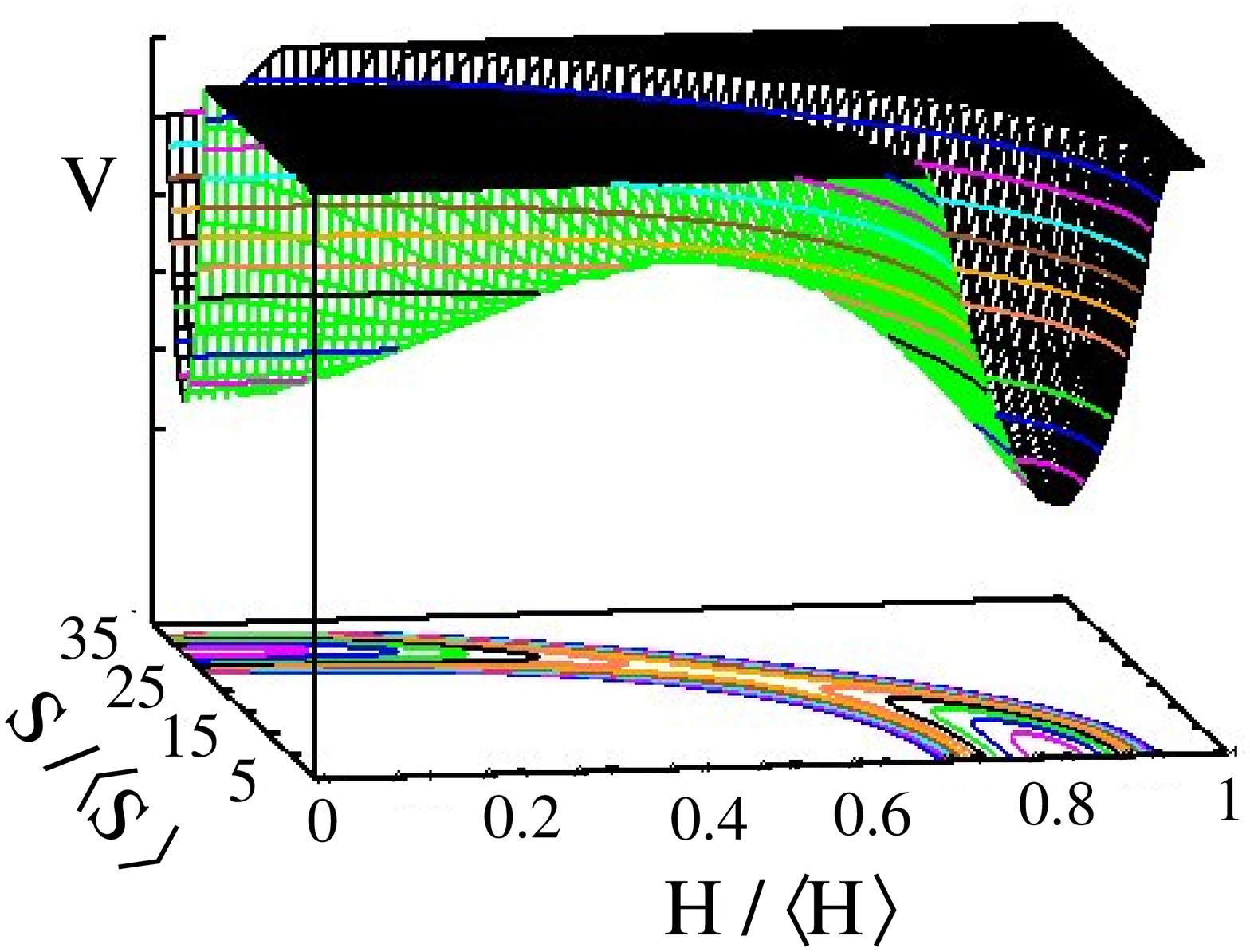}
\epsfxsize=0.5\textwidth\epsfbox{smc-trend2.eps}}
\caption{(a) Left: potential at critical temperature for
typical point having smaller couplings: 
$\lambda_{hs}= 0.633$, $\lambda_h=  0.633$,
$\lambda_s = 0.167$, $y_i^2 =   1.25$, $\langle s\rangle =   10$
GeV. (b) Right: 
Dependences of $T_c$, $v_c$, $v_c/T_c$, $S_{c1}$, and
$S_{c2}$ on $y_i^2$ around the point specified in (a).}
\label{smctrend}
}

It is interesting to know how important it is to have large coupling
constants to get a significant effect on the EWPT. With such large
couplings as the maximum values used in the scan described above,
one expects to reach a Landau pole in one of the couplings
at a relatively low renormalization
scale $\mu_{\rm max}$.  
We have investigated this by integrating the renormalization group
equations for $\lambda_h$, $\lambda_s$, $\lambda_{hs}$,
$y_i$ and $y_t$ (appendix \ref{betafns}).  The distributions of
$\mu_{\rm max}$ plotted in figure \ref{rge} show that indeed new
physics beyond the singlet and the right-handed neutrinos must
typically come in at the scale of several TeV.  
(This would also 
be true
in any model which uses the mechanism of ref.\ \cite{Carena}, since
strong couplings are always needed to get that effect.)  

This motivated us to further explore the model at somewhat weaker
interaction strengths.  We thus made an additional scan of the same point density
as described previously, but in the more limited range
\beq
	0 < \lambda_s,\lambda_h < 1,\quad
	-1 < \lambda_{hs} < 1,\quad
	0 < y_i^2 < 2,\quad 
	\langle s\rangle < 1800  {\rm\ GeV}
\label{ranges-s}
\eeq
Encouragingly, none of the accepted parameters from this range 
fail the Landau pole test, showing that indeed perturbation 
theory is more reliable in this case.  (In fact $\mu_{\rm max}$
is always greater than 50 TeV for this sample.) 
We find that almost no points are accepted for $\lambda_{hs}<0$, 
nor for $\langle s\rangle > 100$ GeV, but there is a sizeable number
of accepted parameters with $\lambda_{hs}>0$ and $\langle s\rangle <
50$ GeV.   These points have small mixing angles,
$\sin^2\theta \lsim 0.02$ (due to the LEP constraint), and small singlet masses, $m_S \lsim 20$
GeV.  Their phase transitions tend to follow a circular arc in the 
$H$-$S$
plane, between the $H$ and $S$ axes, as shown in fig.\
\ref{smctrend}(a).  A curious feature is that $S_{c1}$ and $S_{c2}$
are typically an order of magnitude larger than the small zero-temperature
VEV $\langle S\rangle$, due to $\langle H \rangle$ and hence the
mixing being different at high temperature relative to $T=0$. 
The parameter distributions are shown in figure \ref{smc-dist}.
Unlike in the broader region of parameter space described in the 
previous section,  
here we do not find any correlation like that in eq.\
(\ref{linear-corr}); thus these points give rise to a
 strong phase transition for different reasons than the majority of
those in the large coupling regime.

\FIGURE[t]{
\centerline{\epsfxsize=0.9\textwidth\epsfbox{panel-smc-tev.eps}}
\caption{Distributions of accepted parameters with smaller values of
the coupling constants, and $\lambda_{hs} >0$. 
}
\label{smc-dist}
}

The dependence of the EWPT on the Yukawa coupling $y_i^2$ is due to
the strong influence of $y_i^2$ on the $S$ dynamics, which subsequently affects
the dynamics of $H$ through mixing.   To understand this, we examined
the dependences of various quantities, $T_c$, $v_c$, $v_c/T_c$,
$S_{c1}$, and $S_{c2}$ upon the Majorana Yukawa coupling $y_i^2$.  An
example is shown in figure \ref{smctrend}(b).  There is a notable
rise in $v_c$ associated with the decrease in $S_{c2}$---recall that
this is the value of $\langle S\rangle$ in the EWSB vacuum at
$T=T_c$; see fig.\ \ref{trans}. We can give an analytic explanation
for the relation between $v_c$ and $S_{c2}$.  To this end, it is
useful to think in terms of an effective potential along the light
direction $H'$, which we approximate by the straight line paths
connecting the symmetric and EWSB vacua shown in  figure
\ref{trans}.  At the critical temperature, we can write  \beq
\left(\begin{array}{c} H\\ S\end{array}\right) = 
\left(\begin{array}{c} 0\\ S_{c1}\end{array}\right)
+\left(\begin{array}{cc} $\phantom{-}$\cbt & \sbt \\ -\sbt & \cbt
\end{array} \right) \left(\begin{array}{c} H'\\ S'\end{array}\right)
\eeq where the mixing angle $\bar\theta$ is generally different from
the zero-temperature mixing angle $\theta$.  At $T=T_c$, the shape of
the  potential is roughly of the form $\lambda' H'^2 (H'-v'_c)^2 = 
\lambda H'^4 -2g' H'^3 + \mu_c^2 H'^2$.  From this form, we see that
$v'_c = g'/\lambda'$.  The cubic term can be estimated from the 
tree-level potential as \beq V_{\rm cubic}	=-2\sbt 
\left(\lambda_{hs}\cbt^2 +2\lambda_s\sbt^2\right) S_{c1} 
H'^3\quad\to\quad g' = \sbt  \left(\lambda_{hs}\cbt^2
+2\lambda_s\sbt^2\right) S_{c1} \label{cubicterm} \eeq Similarly, the
effective quartic coupling is $\lambda' = \cbt^4\lambda_h
+2\cbt^2\sbt^2\lambda_{hs} +\sbt^4  \lambda_s$. This gives the
estimate \beq {v_c}\ \sim\ \cbt\,S_{c1}{\left(\sbt\cbt^2\lambda_{hs}+
2\sbt^2\lambda_s\right)\over ( \cbt^4\lambda_h
+2\cbt^2\sbt^2\lambda_{hs} +\sbt^4 \lambda_s) } \label{vcestimate}
\eeq

From eq.\ (\ref{vcestimate}) and fig.\ \ref{trans} it is clear that
for $\lambda_{hs}>0$, a decrease in $S_{c2}$ leads to an increase
the mixing angle $\bar\theta$ and consequently an increase in $v_c$.
This accounts for the initial growth in $v_c$ for $y_i^2\lsim 0.6$
in fig.\ \ref{smctrend}.  Beyond this point, $S_{c2}$ remains
constant, but $S_{c1}$ decreases (hence  $\bar\theta$ decreases),
leading to a decrease in $v_c$.  At the same time, eq.\ (\ref{Tc})
shows that, for $\lambda_{hs}>0$, decreasing $S_{c1}$ leads to an
increase in $T_c$.  Both of the these effects cause $v_c/T_c$ to go
down with $y_i^2$ as observed in fig.\ \ref{smctrend}(b).

\subsection{Constraints from electroweak precision observables}\label{ewpo}

In refs.\ \cite{Profumo:2007wc,Barger:2007im} it was noted that
electroweak precision observables provide a strong constraint 
on the related model containing a real singlet field.  It is known that
the oblique parameters $S,\,T,\,U$ are best fit by a light Higgs boson,
and this preference thus extends to singlets that mix with the doublet
Higgs.  The same constraints as for the real singlet apply to the
Majoron model, since the extra Goldstone boson does not mix and
therefore plays no role.  We have thus carried out the same
analysis as in \cite{Profumo:2007wc,Barger:2007im} to further
constrain the accepted parameter sets described above. 
For completeness, the relevant formulas are given in appendix \ref{EWPOS}.  
As a check on our implementation, we reproduced the results shown in 
figures~9 and 10 of~\cite{Profumo:2007wc}.  

The EWPO constraint indeed has a strong impact on the accepted 
parameter distributions.  In our larger coupling sample, 
730 out of 1000 points are removed for $\lambda_{hs} <0$, and
1650 out of 1710 are excluded for $\lambda_{hs} > 0$;  overall
88\% of otherwise accepted points are thus ruled out, at 95\% c.l.
The resulting distributions are shown in fig.\ \ref{hists-ewpo}.
The most striking difference relative to the corresponding results
without EWPO, fig.\ \ref{hists}, is the elimination of large values
of the doublet-like mass $m_{H'}$, and the restriction to smaller
mixing angles, $\cos^2\theta \gsim 0.8$. There is also a stronger
exclusion of small values of the Majorana Yukawa coupling.

\FIGURE[b]{
\centerline{\epsfxsize=\textwidth\epsfbox{panel2.eps}}
\caption{Distributions of accepted parameters with larger values of
the coupling constants, after EWPO cut.
}
\label{hists-ewpo}
}

In the small-coupling sample, the EWPO constraint is even more 
powerful, eliminating 4370 out of 4860 parameter sets.  Again, the
effect is to eliminate higher values of the Higgs boson
mass.  The distributions are plotted in fig.\ \ref{smc-ewpo}.
The lower $m_{H'}$ values lead to correspondingly smaller values of
the coupling $\lambda_h$.

\FIGURE[t]{
\centerline{\epsfxsize=\textwidth\epsfbox{panel-smc-ewpo.eps}}
\caption{Distributions of accepted parameters with smaller values of
the coupling constants, after EWPO cut.}
\label{smc-ewpo}
}

Two cautionary remarks are, however, in order here.
First, since the additional singlet can be light, a precision analysis 
of EWPO would require the inclusion of further parameters 
$U,\,V,\,X$~\cite{Maksymyk:1993zm,Burgess:1993mg}. 
Here, we do not go beyond $S,T,U$ but leave an extensive EWPO analysis 
for future work. 
Second, the presence of additional new physics beyond the singlet-extension 
of the SM could considerably weaken the EWPO constraints. 
Therefore we refrain from imposing EWPO as a strict constraint in the following.
Rather, we present results both with and without this constraint.

\section{Implications for LHC}
\label{LHCsect}

From the perspective of collider phenomenology, it is quite intriguing 
that all our accepted points feature a relatively light scalar, either the singlet- 
or the doublet-like state, with mass less than about 200 GeV. 
The other state is typically considerably heavier. 
Concerning detectability at the LHC, it is important to to know how the 
mass of the lighter state correlates with the mixing angle, 
{\it i.e.,} 
how its couplings compare to those of a SM Higgs boson.
To this purpose,  figure~\ref{lepdist} shows the scale factor $\xi$ 
of the squared couplings versus mass of the lighter scalar.  Specifically, 
$\xi=\cos^2\theta$ for the Higgs-like boson, while $\xi=\sin^2\theta$ for 
the singlet-like state.  
This can be
compared with the LEP bound (fig.\ \ref{lepfig}) to get a feeling for
how easily detectable the light boson may be at the LHC. 

\FIGURE[t]{\centerline{
\epsfxsize=0.5\textwidth\epsfbox{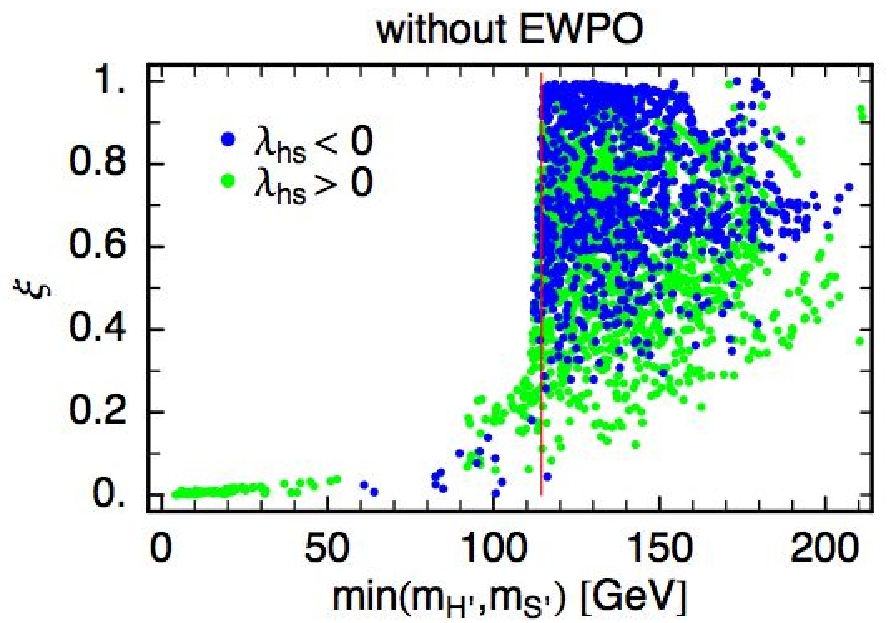}
\epsfxsize=0.5\textwidth\epsfbox{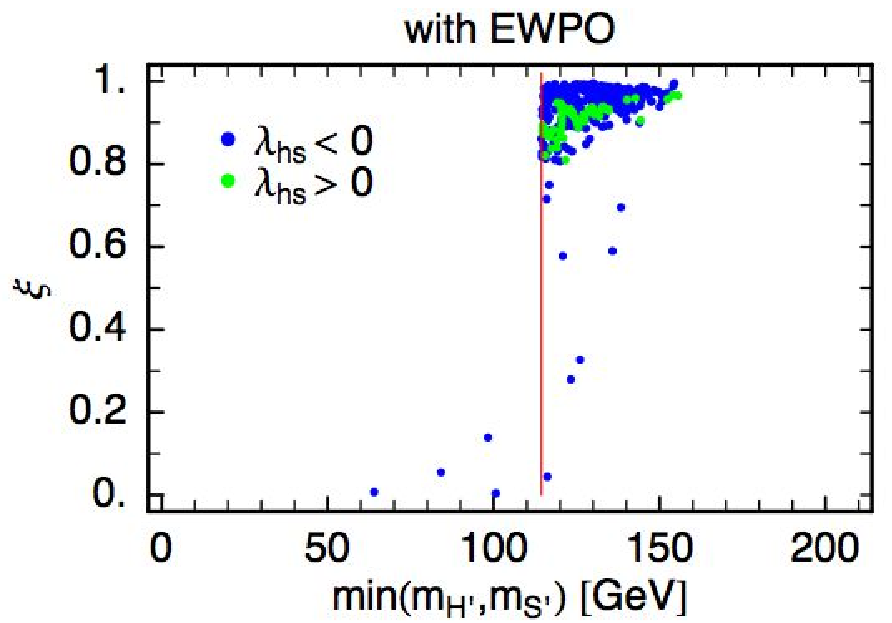}}
\vspace*{-6mm}
\caption{Distribution of scale factor of the squared couplings (relative to a SM Higgs) 
versus mass of the lighter scalar. For a doublet-like state
$\xi=\cos^2\theta>0.5$, while for a singlet-like state 
$\xi=\sin^2\theta<0.5$. The vertical red line indicates the limit on a 
SM Higgs boson. The left (right) panel is without (with) the EWPO constraint.}
\label{lepdist}
}

Let us first discuss the situation without the EWPO constraint, 
shown in the left plot of figure~\ref{lepdist}. The density
of points in this scatter plot indicates that there would have been
many examples providing a strong EWPT in the LEP-excluded region,
but there is not a strong bias toward being close to the limit.
Nevertheless, there is an upper limit of 
${\rm min}(m_{H'},\,m_{S'})\lsim 200$~GeV 
from the requirement of a strong EWPT 
(note also the lower limit on the mostly-doublet state of 
$m_{H'}\gsim 113$~GeV from LEP data).
Moreover, a large fraction of the accepted points features a sizable 
doublet--singlet mixing. 

In the sample with small values of the couplings, the situation is
different, because only small values of $m_S \lsim 20$ GeV, and the 
mixing angle $\sin^2\theta\lsim 0.02$, are present.  This situation was considered for
the similar model of a real singlet field in ref.\ \cite{lightS}. 
There it was noted that values of $m_{S'} \lsim 5$ GeV (the $B$ meson
mass) are strongly constrained ($\theta<10^{-2}$) by the decays $B\to
S' X$ followed by $S'\to \mu^+\mu^-$.  For $m_{S'}$ in the range 
5 GeV $\lsim m_{S'} \lsim $ 50 GeV, the Higgs can decay into singlets, $H'\to S'S'$ at a level which
can compete with the two photon final state, $H'\to \gamma\gamma$. For
the accepted parameters in the small coupling regime, we find that
the right-handed neutrinos are always heavier than $m_{S'}/2$. 
Therefore there is never an invisible decay channel $S'\to\nu_R\nu_R$
in this case.  Instead, $S'$ decays predominantly into $b\bar b$ quark
pairs due to the small doublet--singlet mixing.

The situation changes quite drastically when applying the EWPO constraint, 
as shown in the right plot of figure~\ref{lepdist}. In this case, the allowed 
range shrinks to ${\rm min}(m_{H'},\,m_{S'})\lsim 156$~GeV, and the regions 
with large mixing and/or a light singlet are almost completely cut away.  

\FIGURE[t]{\centerline{
\epsfxsize=0.5\textwidth\epsfbox{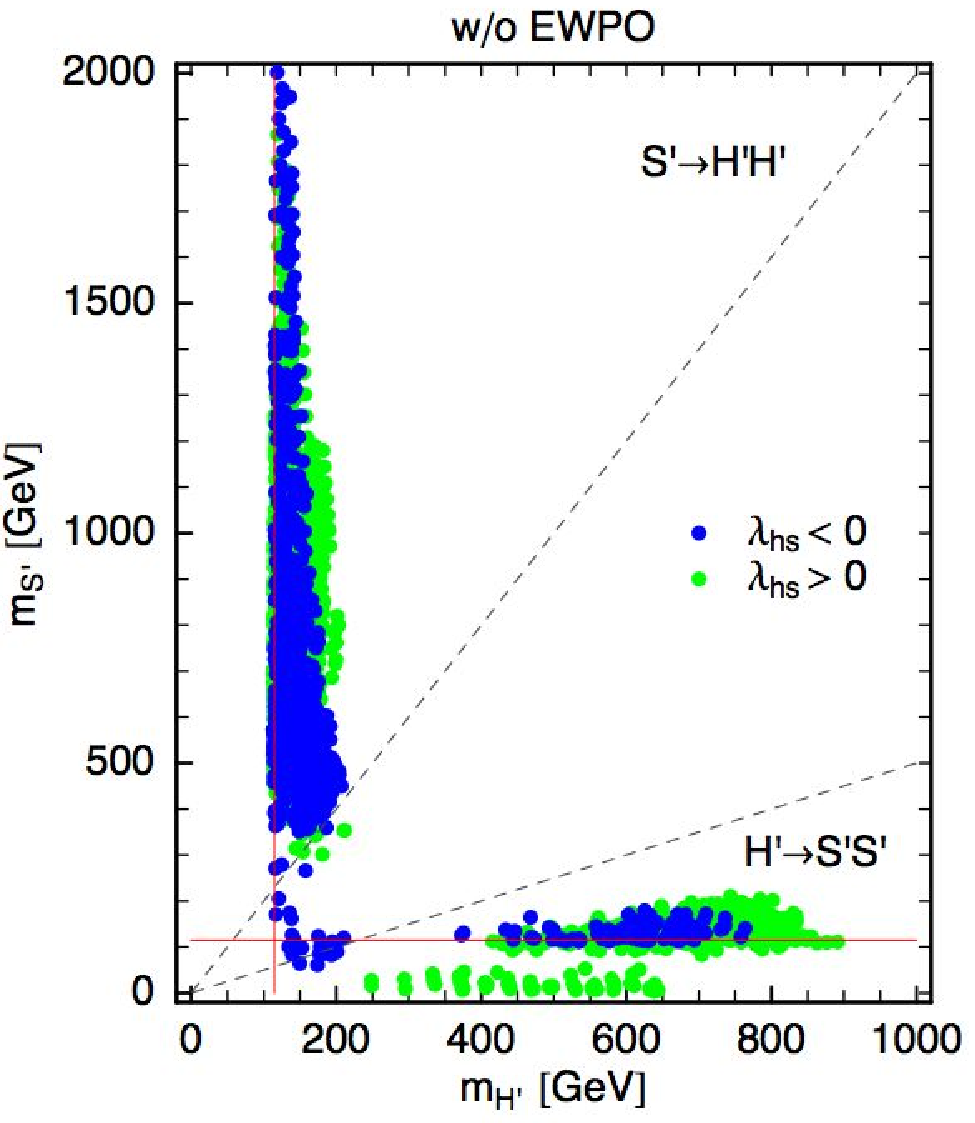}
\epsfxsize=0.5\textwidth\epsfbox{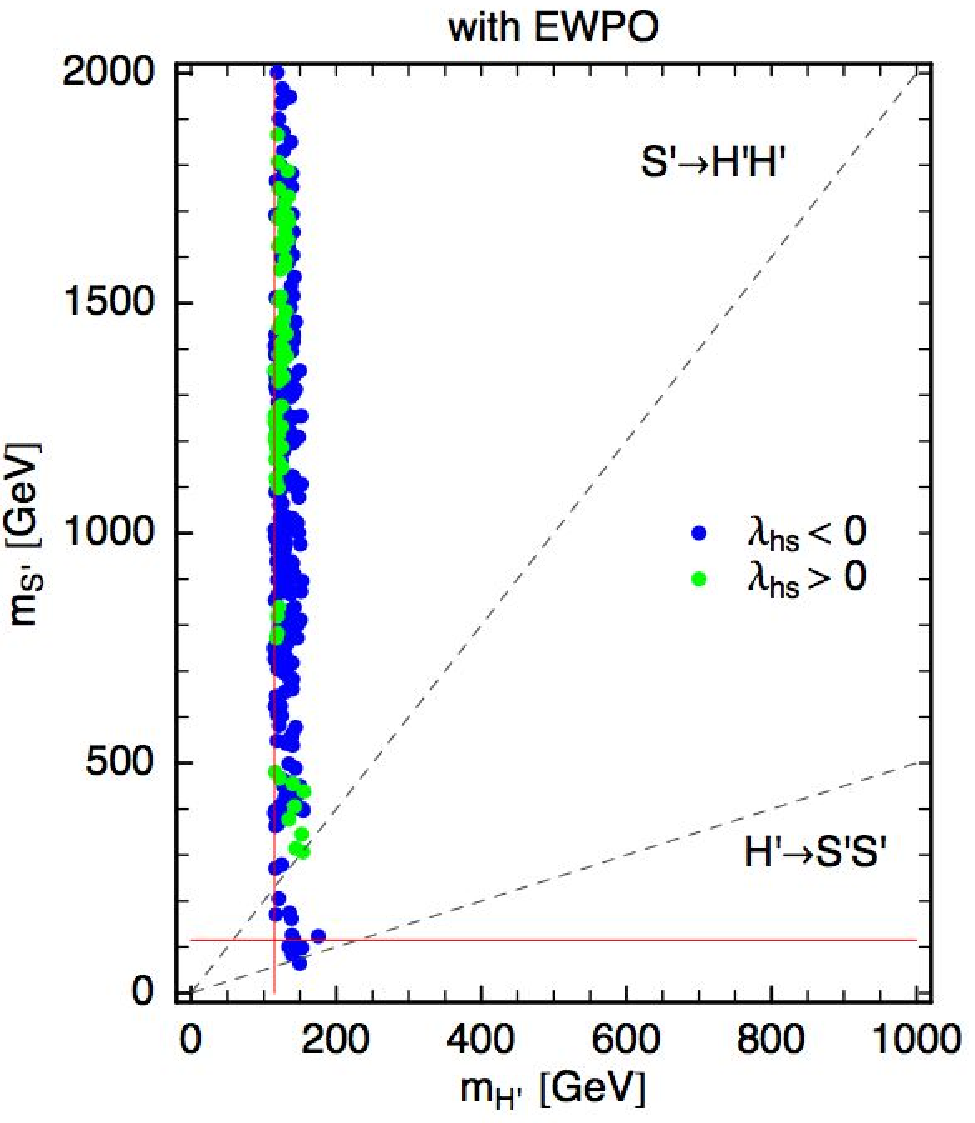}}
\vspace*{-1.6cm}
\caption{Scatter plots of the mostly-singlet versus mostly-doublet
Higgs masses without (left) and with (right) EWPO constraint. 
The full red lines (horizontal and vertical) indicate the SM limit of $m_H^{\rm SM}>114.4$~GeV.
Above the upper dashed lines $S'\to H'H'$,  
below the lower dashed lines $H'\to S'S'$ is kinematically allowed.}
\label{mcorr}
}

It is also useful to consider the correlation between the mostly-singlet
and mostly-doublet Higgs masses, shown in figure~\ref{mcorr}, 
which reveals that typically one state is significantly lighter than the other. 
For both, $\lambda_{hs}<0$ and $\lambda_{hs}>0$, we see distinct islands 
with either (i) $m_{H'}\lsim200$~GeV and  $m_{S'}$ 
ranging from few hundred GeV up to order TeV, or 
(ii) $m_{H'}>200$~GeV and a (much) lighter singlet.  
The latter region is, however, completely removed by the 
EWPO constraint.  
For $\lambda_{hs}<0$, there also exists a small region with 
$m_{H'}\sim115-200$~GeV and $m_{S'}\sim50-130$~GeV.  
The triple correlation between the two masses and the mixing angle is shown
in figs.~\ref{tcorr} and \ref{tcorr2}.  

\FIGURE[t]{\centerline{
\epsfxsize=0.5\textwidth\epsfbox{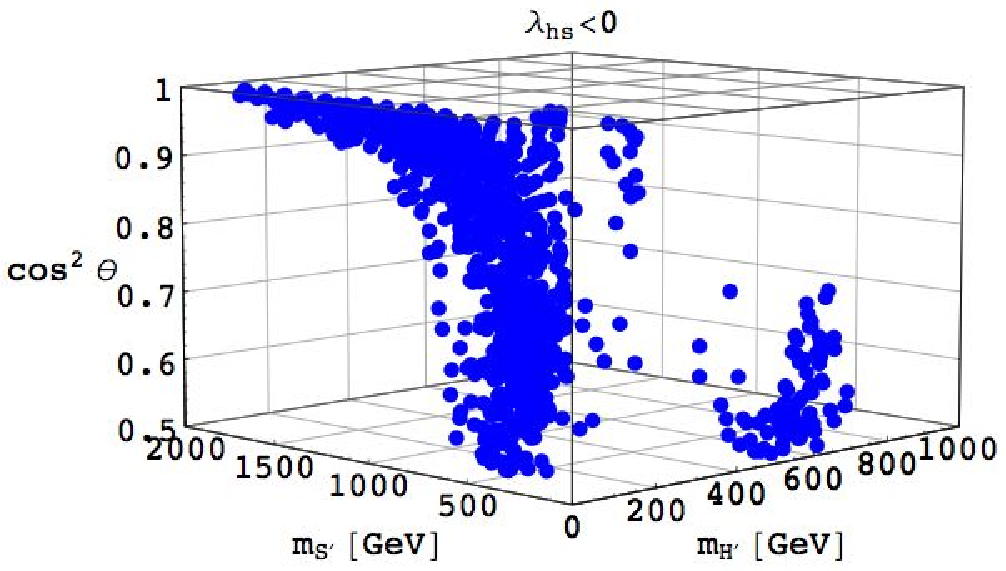}
\epsfxsize=0.5\textwidth\epsfbox{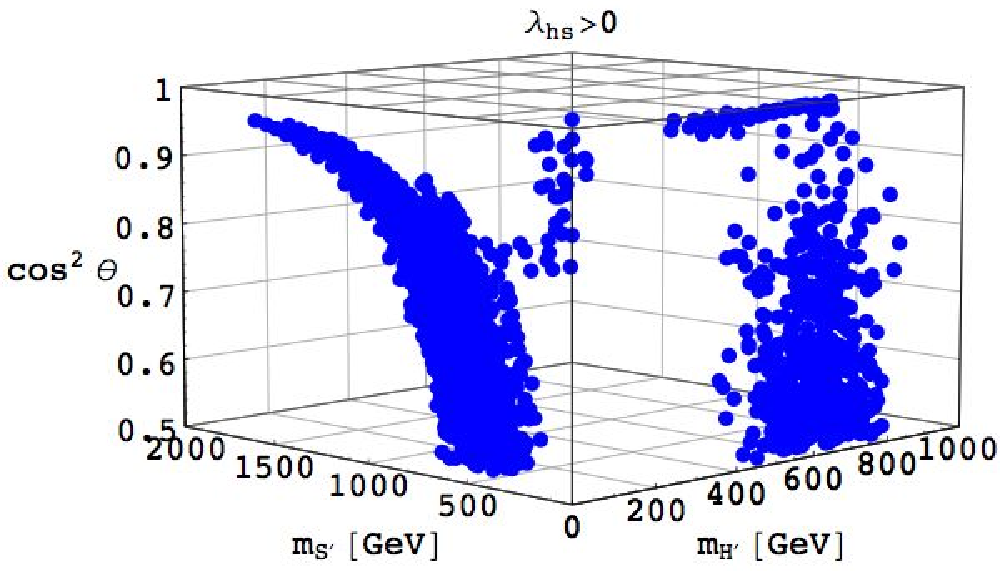}}
\vspace*{-8mm}
\caption{Triple correlation of mass eigenvalues and mixing angle
for $\lambda_{hs}<0$ (left) and $\lambda_{hs}>0$ (right); without EWPO 
constraint.}
\label{tcorr}
}
\FIGURE[t]{\centerline{
\epsfxsize=0.5\textwidth\epsfbox{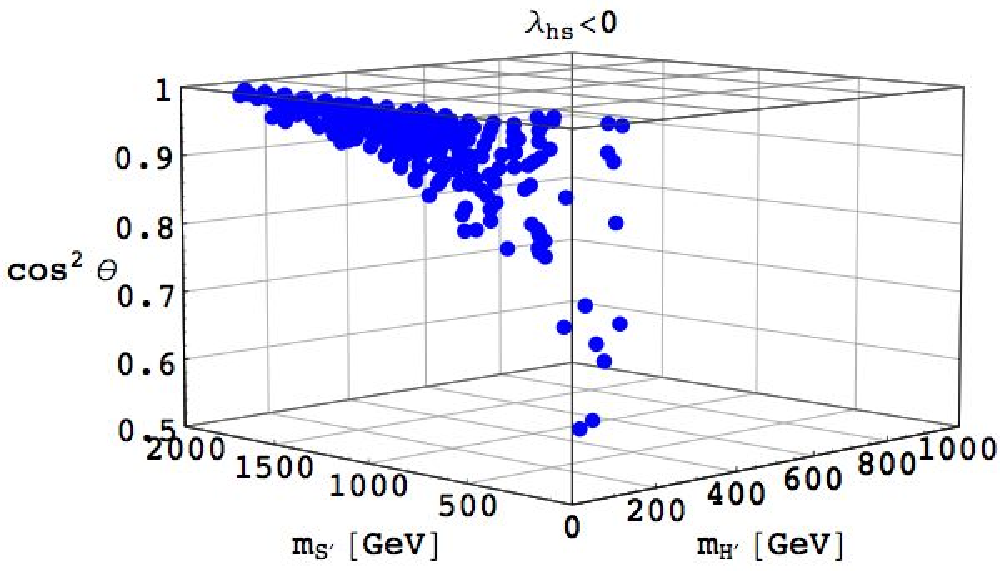}
\epsfxsize=0.5\textwidth\epsfbox{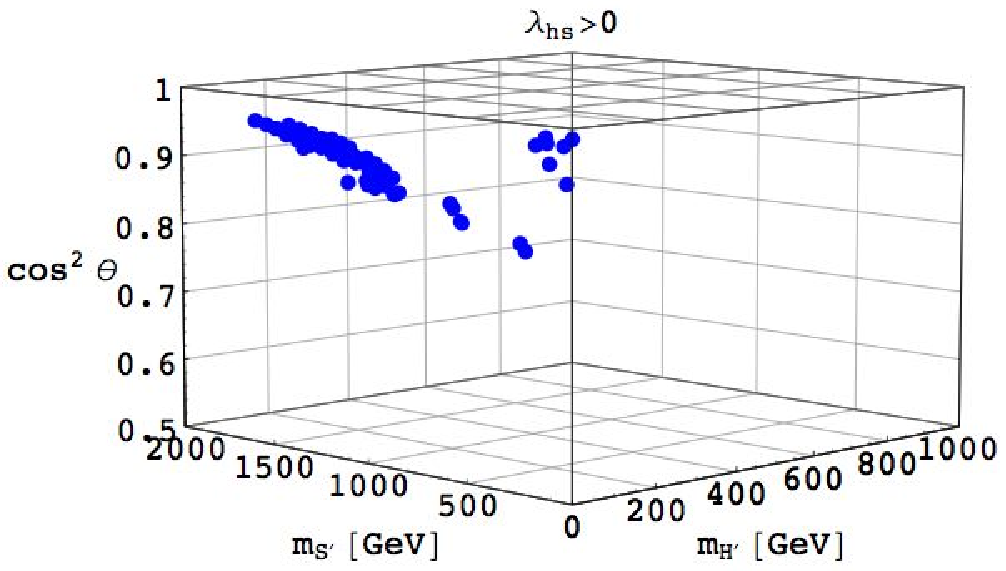}}
\vspace*{-8mm}
\caption{Same as figure~\ref{tcorr} but with EWPO constraint applied.}
\label{tcorr2}
}

Regarding the decay modes, we see from figure~\ref{mcorr}  that for
most points Higgs-to-Higgs decays, either $S'\to H'H'$ or $H'\to
S'S'$,  are allowed, which could modify the branching ratios relative
to those of a SM Higgs boson.  (If Higgs-to-Higgs decays are absent,
the branching ratios of both the $H'$ and  the $S'$ are just the same
as those of a SM Higgs.) Denoting the lighter of the two states
as $h_1$ and the heavier one as $h_2$, the generic 
expression for the decay width is  
\beq
  \Gamma(h_2\to h_1h_1) = \frac{g_{211}^2}{8\pi m_2}\sqrt{ 1-4m_1^2/m_2^2 }\,,
\eeq
where cubic coupling $g_{211}$ is given by
\beq
	g_{H'S'S'} = 24\lambda_h\langle h\rangle c_\theta^{} s_\theta^2 - 
	24\lambda_s\langle s\rangle s_\theta^{} c_\theta^2
     	+4\lambda_{hs}\left(\langle h\rangle c_\theta^{}(c_\theta^2-2s_\theta^2)-\langle 
	s\rangle s_\theta^{}(s_\theta^2-2c_\theta^2)\right)\, 
\eeq
for $h_2=H'$, $h_1=S'$ and
\beq	
        g_{S'H'H'} = 24\lambda_h\langle h\rangle s_\theta^{} c_\theta^2 + 
	24\lambda_s\langle s\rangle c_\theta s_\theta^2
     	+4\lambda_{hs}\left(\langle h\rangle s_\theta^{}(s_\theta^2-2c_\theta^2)+
	\langle s\rangle c_\theta^{}(c_\theta^2-2s_\theta^2)\right)
\eeq
for $h_2=S'$, $h_1=H'$. Here, $c_\theta=\cos\theta$ and $s_\theta=\sin\theta$. 
The branching ratio is then given by 
\beq 
  BR(h_2\to h_1h_1) = 
  \frac{\Gamma(h_2\to h_1h_1)}{\xi \Gamma(H_{\rm SM})+\Gamma(h_2\to h_1h_1)}\,,
\eeq
where $\xi=\cos^2\theta$ ($\sin^2\theta$) for $h_2=H'\,(S')$ and  
$\Gamma(H_{\rm SM})$ is the total decay width of the SM Higgs boson 
with same mass as $h_2$. We use {\tt HDECAY} \cite{Djouadi:1997yw} to compute 
$\Gamma(H_{\rm SM})$. 
In the numerical analysis, we find that $S'\to H'H'$ decay typically has a rate 
of only a few percent, $BR(S'\to H'H')\lsim 5\%$, and is therefore negligible  over most of the parameter
space.\footnote{The branching ratio of $S'\to H'H'$  is enhanced for
$\cos\theta\to 1$, but in this case the $S'$ production rate goes to
zero.  Our data set contains one point with $m_{S'}=452$~GeV, 
$m_{H'}=173$~GeV, $\cos\theta=1$ and BR$(S'\to H'H')=100\%$.  All
other points have $BR(S'\to H'H')\lsim 5\%$.}
For the mostly-doublet state, on the other hand, Higgs-to-Higgs
decays  can be important in the region $m_{H'}< 200$~GeV and
$m_{S'}\sim 50-100$~GeV.   Our data set contains six points with
$m_{H'}\sim 175-200$~GeV and 
$m_{S'}\sim 60-95$~GeV which have $BR(H'\to S'S')\sim 3-8\%$. 
With $\cos\theta\gsim 0.9$ these points have small doublet-singlet
mixing. The singlet here decays 80--85\%
of the time to $b\bar b$ and about 9\% to $\tau^+\tau^-$,
leading to $4b$, $2b2\tau$ and $4\tau$ final states. 
We also find one point with $m_{H'}=150$~GeV, 
$m_{S'}=64$~GeV, $\cos\theta=0.99$ and BR$(H'\to S'S')=56\%$. 
This point remains after EWPO contraints.  It is also worthwhile to
remember that, even when Higgs-to-Higgs decays are absent or negligible, 
the total decay width is modified by a factor $xi=\cos^2\theta$
($\sin^2\theta$) in the case of  $H'$ ($S'$), relative to the SM Higgs
boson.

Let us finally discuss the discovery potential at the LHC. 
To this end we use the CMS expectations on SM Higgs boson searches 
presented in \cite{Ball:2007zza}. Figure 10.38 of \cite{Ball:2007zza} shows 
the luminosity needed for a $5\sigma$ discovery in various standard 
search channels. For $H'$ and $S'$, this luminosity scales with $1/\xi^2$ 
due to the reduced production cross section, 
and, where applicable, a factor stemming from the modification of 
the branching ratios into SM particles; for the $H'$:
\beq
   \frac{ BR(H'\to X_{\rm SM}) }{ BR(H_{\rm SM}\to X_{\rm SM}) } = 
   \frac{\xi\,\Gamma(H_{\rm SM})}{\xi \Gamma(H_{\rm SM})+\Gamma(H'\to S'S')}\,,
\eeq
with $\xi=\cos^2\theta$, and analogously for $S'$ with $H'\leftrightarrow S'$ 
and $\xi=\sin^2\theta$.  
The resulting luminosity needed for a $5\sigma$ discovery is shown in 
figure~\ref{lumi}; the left (right) plot shows the lighter (heavier) 
mass eigenstate. Blue dots represent a mostly-doublet Higgs, 
green dots a mostly-singlet one; 
points in darker colour are those which survive the EWPO constraints. 
The number and nature of Higgs bosons which are within discovery reach 
with $30\,{\rm fb^{-1}}$ of data is shown in figure~\ref{reach}.

\FIGURE[t]{\centerline{
\epsfxsize=0.5\textwidth\epsfbox{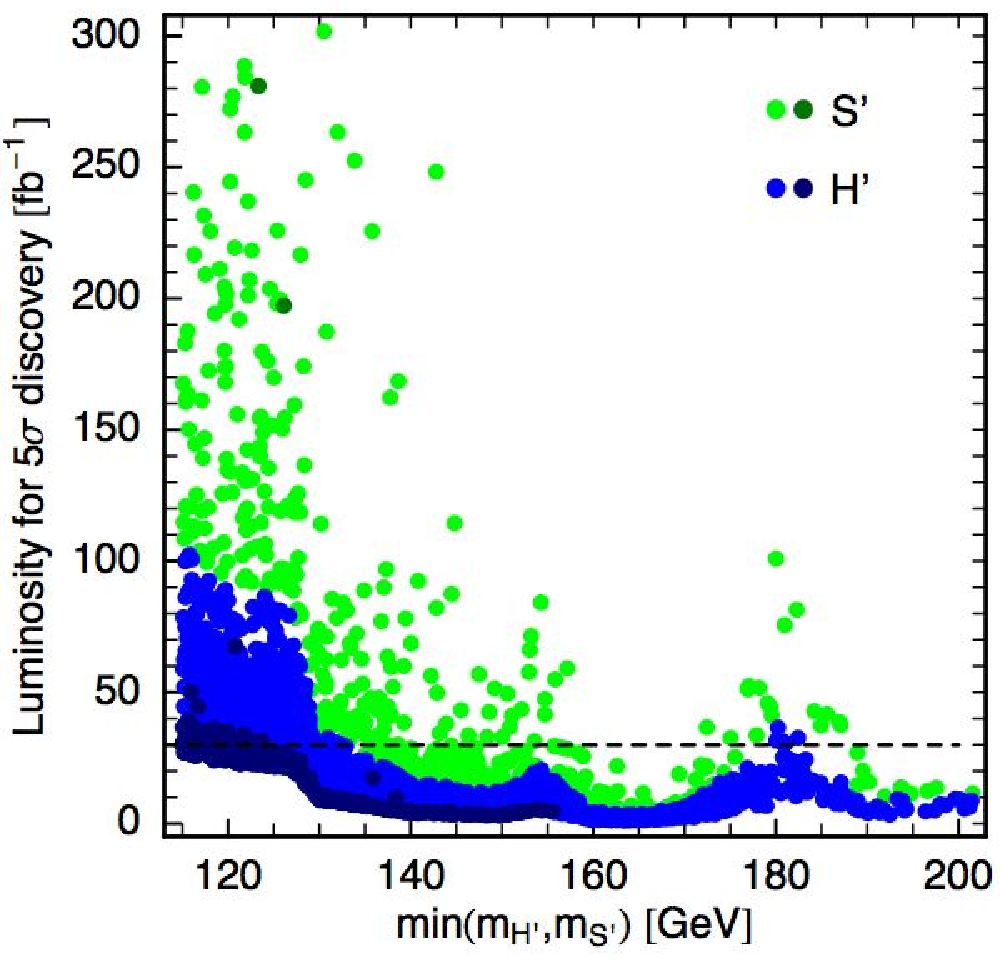}
\epsfxsize=0.5\textwidth\epsfbox{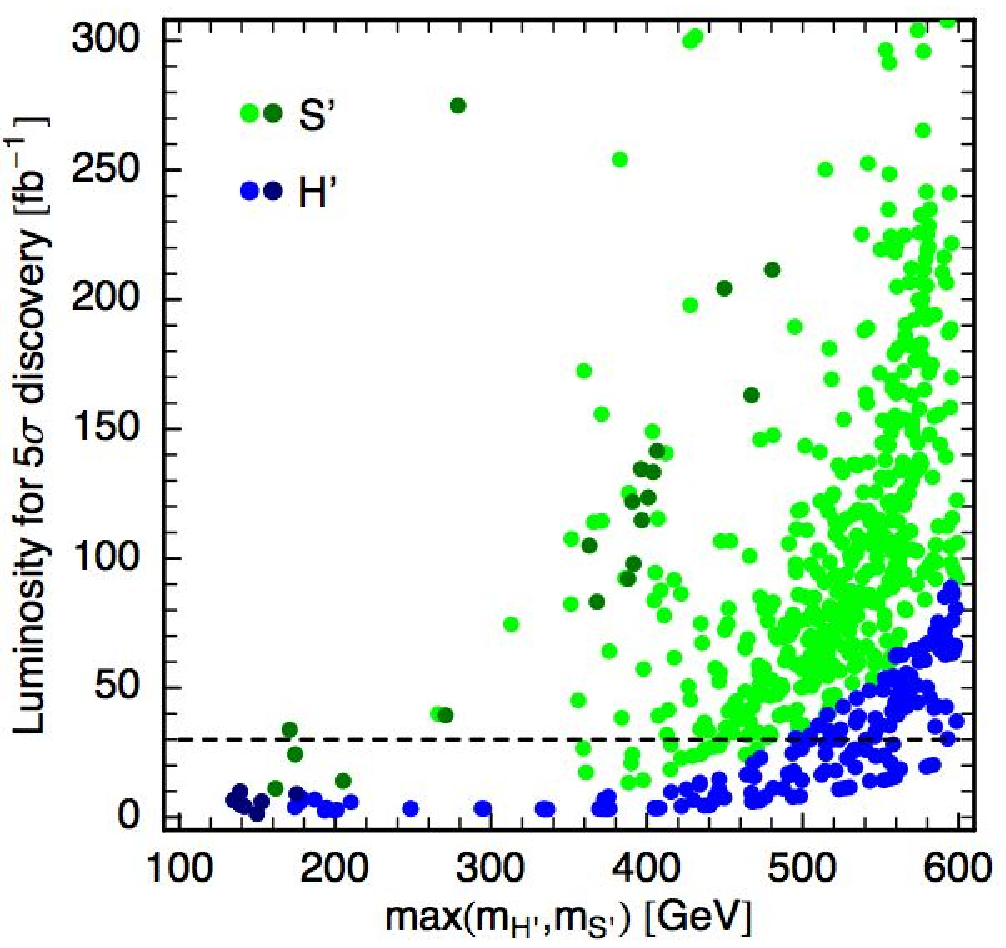}}
\vspace*{-8mm}
\caption{Luminosity needed for a $5\sigma$ discovery of the lighter (left plot) and 
the heavier (right plot) mass eigenstate, extrapolated from CMS results~\cite{Ball:2007zza}. 
Blue dots represent a mostly-doublet Higgs, green dots a mostly-singlet one; 
points in darker colour survive EWPO constraints. 
The horizontal dashed lines indicate $30\,{\rm fb^{-1}}$ or 
three years of running at 
low luminosity.}
\label{lumi}
}

\FIGURE[t]{\centerline{
\epsfxsize=0.5\textwidth\epsfbox{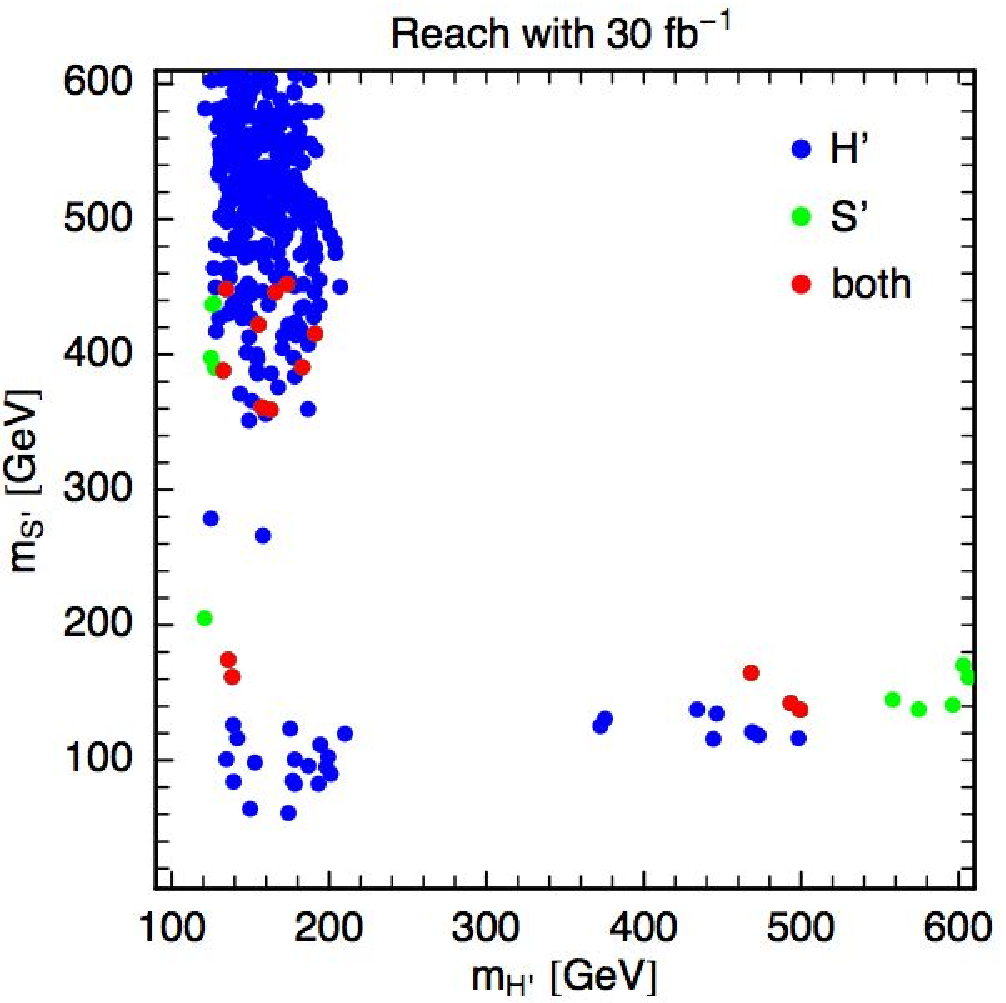}
\epsfxsize=0.5\textwidth\epsfbox{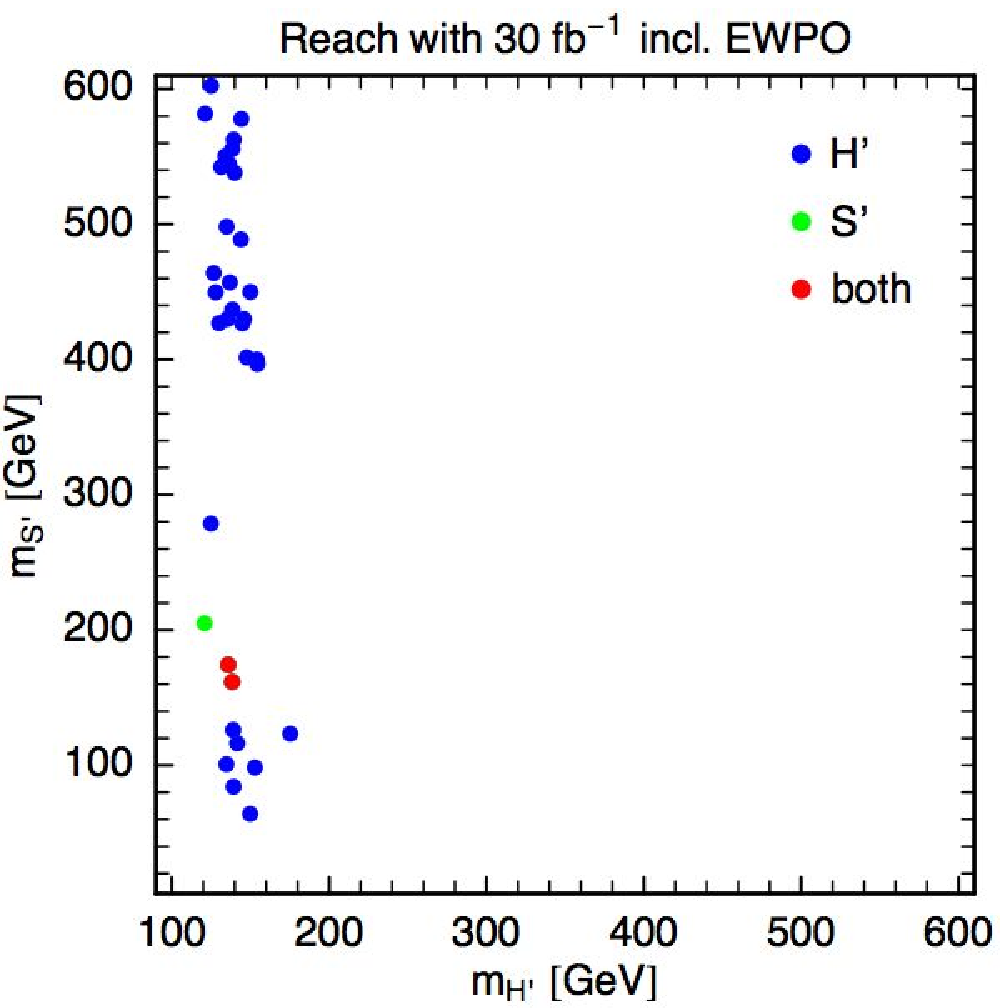}
}
\vspace*{-8mm}
\caption{Discovery reach with $30\,{\rm fb^{-1}}$ in the $m_{S'}$ versus $m_{H'}$ plane. 
The different colours encode which state can be detected: blue stands for $H'$, green 
for $S'$, red for both. The left plot is without, the right one with EWPO constraints.}
\label{reach}
}
 
In summary, we conclude that there is a strong discovery potential for this
nonstandard Higgs sector, if it is the origin of a strong EWPT. 
This holds in particular for the mostly-doublet Higgs, but in many cases also 
for the mostly-singlet one. There are even a couple of points where both 
$H'$ and $S'$ could be discovered at the LHC with $30\,{\rm fb^{-1}}$
of integrated luminosity. 
With higher luminosity, the prospects for discovering both mass eigenstates
are quite promising. The alert reader will note, however, that figure~\ref{lumi} 
is limited to $115~{\rm GeV}\le m_{H',S'}\le600~{\rm GeV}$. In figure~\ref{reach}, 
we also show $m_{S'}<115$~GeV, but  the $S'$ can be observed in this
case.
The reason is that the experimental analyses in \cite{Ball:2007zza} only cover the 
mass range $115~{\rm GeV}\le m_{H}\le600~{\rm GeV}$. In the majoron model, 
as well as in other singlet-extensions of the SM, it would however be interesting 
to search also for lighter and heavier states. 
We hope that the present work provides some motivation toward this end.

\section{Conclusions}
\label{conclusions}

We have given an in-depth analysis of the EWPT in the singlet Majoron
model at the one-loop level, taking account of the LEP constraints on
the Higgs boson mass and mixing angle.  Our broad scan of the model's
parameter space indicates that a certain level of tuning is needed to
get a strong enough transition for electroweak baryogenesis; many
examples approximately satisfy the relation (\ref{linear-corr}) which
reduces the dimensionality of the parameter space.   The
zero-temperature one-loop correction was shown to play a crucial role
in this respect, effectively reducing the negative mass squared of 
the Higgs and hence the critical temperature of the phase transition,
to increase the figure of merit for the strength of the transition,
$v_c/T_c$.   These examples required rather large values of the
coupling constants, as large as $\lambda_i\lsim 3$.  
We also identified another population of accepted points with
$\lambda_i\lsim 1$ which has a different origin, and exists only for
very small singlet masses $m_S\lsim 20$ GeV.  

From the technical point of view, our job of identifying cases with a
first order phase transition was made more difficult by the fact that
both fields $H$ and $S$ usually evolve during the transition, so it
is essential to keep track of both.  Although numerous studies have
been done on similar models with a singlet coupling to the Higgs,
most of these assume that the singlet does not get a VEV, so this
complication does not arise.  Another difference is that a generic
model of a real singlet interacting with $H$ has many additional
couplings which are odd in $S$, whereas the singlet Majoron model is
constrained by the $U(1)$ global lepton symmetry, which it
spontaneously breaks.  Due to the reduced number of coupling
constants, we were able to make an exhaustive search of the parameter
space.   The singlet Majoron model also has the appeal of being
theoretically motivated by the seesaw mechanism for neutrino masses.
In our case, this must be supplemented by the requirement of small 
Dirac Yukawa couplings for the neutrinos, since we take the
right-handed neutrinos to be lighter than the TeV scale. 

Of course the strong phase transition is only interesting for
baryogenesis if there is also a mechanism for producing the baryon
asymmetry.  Complex Majorana Yukawa couplings, which we assumed here
to be real for simplicity, could provide the needed CP violation. 
Perhaps CP-violating reflections of the heavy neutrinos at the bubble
walls could create a lepton asymmetry which would be converted to the
baryon asymmetry via sphaleron interactions.   This is a subject to
which we hope to return.  

It would be interesting to extend our study to the generation of 
gravitational waves.  Although electroweak baryogenesis and gravity
wave generation both need a ``strong'' phase transition, the criteria
are different.  In particular, relativistic bubble walls are favored
for producing significant gravity waves.  Ref.\ \cite{guy} has
recently shown that this can be achieved in the related model of a
real singlet with more general couplings (not respecting any $Z_2$
symmetry) than in the Majoron model.   

Concerning collider phenomenology, we have shown that the LHC 
has a strong discovery potential for this nonstandard Higgs sector, 
if it is the origin of a strong EWPT. This includes the possibility of a 
SM-like Higgs boson with mass up to about 200~GeV, which has a 
sizeable branching fraction into a pair of light singlets. 
Moreover, with high enough luminosity 
there are good prospects to discover both the $H'$ 
and the $S'$ states. A dedicated experimental study would be 
worthwhile to cover masses below 115~GeV and above 600~GeV.

\acknowledgments
We thank H.\ Logan for helpful correspondence, 
G.\ Moore for valuable comments and suggestions, and 
M.\ Ramsey-Musolf for his notes and patient discussions about EWPO
constraints.  
We also thank Alexandre Nikitenko for providing explicit numbers 
for the CMS results. 
JC is supported by the Natural Sciences and Engineering Research
Council of Canada. The work of SK is supported by the French ANR
project ToolsDMColl, BLAN07-2-194882.

\appendix

\section{Analytic approximation for thermal potential}
\label{appA}
The method of smoothly matching the low- and high-$T$ expansions
for the one-loop thermal potential was given in ref.\ \cite{cl}.
For convenience we repeat the formulas here.  The $n$th order
high-$T$ (small $M/T$) expansion is given by \cite{AE}
\beqa
&V_{\rm s,b}(n) = -\frac{\pi^2 T^4}{90} + 
\frac{M^2 T^2}{24} - \frac{M^3 T}{12\pi} - \frac{M^4}{64 \pi^2}
\left(\log\left(\frac{M^2}{T^2}\right) - c_b\right)& \nonumber\\
&+ \frac{M^2 T^2}{2}{\displaystyle\sum^n_{l=2}} 
\left(\frac{-M^2}{4\pi^2 T^2}\right)^l
\frac{(2l-3)!! \zeta(2l-1)}{(2l)!!(l+1)}, &{\rm bosons};\nonumber\\
&V_{\rm s,f}(n) = -\frac{7\pi^2 T^4}{720}+\frac{M^2 T^2}{48}  + 
\frac{M^4}{64 \pi^2}
\left(\log\left(\frac{M^2}{T^2}\right) - c_f\right)\nonumber\\
&- \frac{M^2 T^2}{2}{\displaystyle\sum^n_{l=2}}
 \left(\frac{-M^2}{4\pi^2 T^2}\right)^l
\frac{(2l-3)!! \zeta(2l-1)}{(2l)!!(l+1)}\left(2^{2l-1} - 1\right)
,  &{\rm fermions};\nonumber\\
&c_b = 3/2 + 2\log 4\pi - 2\gamma_{\sss E} \cong 5.40762;\qquad
c_f = c_b -2\log4 \cong 2.63503& 
\label{small}
\eeqa	
respectively for bosons and fermions.   The corresponding
low-$T$ (large $M/T$) expansion is \cite{AH}
\beq
V_{\rm l}(n) = 	-  e^{-M/T}\left(\frac{ M T }{2\pi}\right)^{3/2} T
\sum_{l=0}^n \frac{1}{2^l l!}\,\frac{\Gamma(5/2+l)}{\Gamma(5/2-l)}
(T/M)^l.
\label{large}
\eeq

By trial and error, one can find that the low- and high-$T$ expansions
can be smoothly matched onto each other using the approximation
\beqa 
V_b = 
 & \Theta(x_b-(M/T)^2)\, V_{\rm s,b}(3) + 
	\Theta((M/T)^2-x_b)\,(V_{\rm l}(3)-\delta_b\, T^4) 
	& {\rm}\nonumber\\
V_f = & 	\Theta(x_f-(M/T)^2)\, V_{\rm s,f}(4) + 
	\Theta((M/T)^2-x_f)\, (V_{\rm l}(3)- \delta_f\, T^4)
	& {\rm }\label{matching}
\eeqa
where $\Theta$ is the step function with $x_b = 9.47134$ and $x_f = 
5.47281$ for bosons
and fermions, respectively.  The small constant shifts of $V_{\rm l}(3)$
are made so that the function as well as its derivatives match at the
transition point: $\delta_b = 3.19310\times 10^{-4}$ and $\delta_f = 
4.60156\times 10^{-4}$.  This gives an approximation with a relative error which is less
than 0.5\% for $M/T \to \infty$, and negligible for small $M/T$.

\section{Field dependent masses}
\label{appB}

Although it is convenient to express the potential and the
field-dependent masses in terms of the complex VEV's, it is simpler to compute the masses in the real basis, where
$H^0 = (h + i\phi_1)/\sqrt{2}$, $H^+ = (\phi_2 + i\phi_3)/\sqrt{2}$,
$S = (s + ij)/\sqrt{2}$.  We continue to express the field-dependent
masses in terms of the
complex fields, but assuming only $h$ and $s$ actually get VEV's:
$H = h/\sqrt{2}$, $S = s/\sqrt{2}$.  

The diagonal components of the zero-temperature
scalar mass matrix, in the real basis $h,\phi_i,s,j$, are
\beqa
m^2_{h,h} &=&  \lambda_h\left(6|H|^2 - v_h^2 \right)+
	\lambda_{hs}|S|^2\nonumber\\
m^2_{\phi_i,\phi_i} &=&  \lambda_h\left(2|H|^2  -  v_h^2 \right)+
	\lambda_{hs}|S|^2\nonumber\\
m^2_{s,s} &=& \lambda_s\left(6|S|^2- v_s^2\right) + \lambda_{hs}|H|^2
	\nonumber\\
m^2_{j,j} &=& \lambda_s\left(2|S_2|^2-v_s^2\right) + \lambda_{hs}|H|^2
\label{masses}
\eeqa
Within the full $6\times 6$ mass matrix, there is only one
off-diagonal entry, 
\beq
	m^2_{s,h} = m^2_{h,s} =  2\lambda_{hs} |H||S|
\eeq
(Note that we can take $H$ and $S$ to be real here.)  Thus
one can analytically find all the field-dependent mass squared eigenvalues,
by diagonalizing the $h$-$s$ sector.  

For the ring improvement, we must add thermal corrections to the
mass squared matrix,
\beqa
	\delta m^2_{H} &=& T^2\left(\frac12\lambda_h +
\frac1{12}\lambda_{hs}
	+ \frac1{16}(3g^2 + g'^2) + \frac14 y_t^2\right)\\
\label{thermass1}
	\delta m^2_{S} &=& T^2\left(\frac13\lambda_s +
\frac1{6}\lambda_{hs}
	+ \frac{1}{24}\sum_i y_i^2\right)
\label{thermass2}
\eeqa
They can be computed by inserting the zero-temperature masses into
the high-$T$ expansion of the one-loop thermal potential, and reading
off the corrections to the mass terms (the coefficients of the terms
quadratic in $H$ and $S$).  
These thermal masses are the same for each real component ($h,\phi_i$
or $s,j$) within the $H$ or $S$ fields, respectively.  Since 
$\delta m^2_{H} \neq \delta m^2_{S}$, the thermal mass matrix has to 
be diagonalized independently of the zero-temperature mass matrix.

The field-dependent masses of the relevant fermions are given by
\beq
	m^2_t = y_t^2|H|^2,\quad m^2_{\nu_i} = y_i^2 |S|^2
\eeq
They do not need to be thermally corrected for the ring improvement.
In the imaginary time formalism of finite-temperature field theory,
where the effective squared masses of the Matsubara modes are
$M^2(\phi,T) + (2\pi n T)^2$ for bosons and $M^2(\phi,T) + (2\pi
(n+\frac12 T)^2$ for fermions.  Only for the $n=0$ modes of the bosons
can there be an infrared divergence due to vanishing $M^2(\phi)$ which
would make it important to include the perturbative $g^2 T^2$
contribution to $M^2$.

The only other fields we must consider are the gauge bosons.
In the basis of $W_1,W_2,W_3,B$, the mass matrix is
\beq
        m^2_{\rm gauge}(H, T)={|H^2|\over 2}\left(\begin{array}{cccc}
        g^2 & & & \\  & g^2 & & \\ & & g^2 & g g' \\ & & g g' & g'^2 
        \end{array}\right) + T^2\left(\begin{array}{cccc}
        g^2 & & & \\  & g^2 & & \\ & & g^2 &  \\ & & &  g'^2 
        \end{array}\right)
        \left\{\begin{array}{rl}2, & {\rm longitudinal;}\\ 
        0, & {\rm transverse.}\end{array}\right.
\label{gauge}
\eeq
Only the longitudinal components get a thermal correction at leading
order in the gauge couplings.

\section{Renormalization constants}
\label{appC}

To express the solutions to the renormalization conditions
(\ref{renconds}) it is convenient to define  multiplicities $g_i$ for
the respective fields as: 1 for each real scalar, $-12$ for the top
quark, $-2$ for each of the three right-handed neutrinos,  2 for each
transverse gauge boson, 1 for longitudinal gauge bosons.  It is
straightforward to show that
\beqa
\label{lnmu2}
	\ln\mu^2 &=& {\sum_i g_i m^2_i {\partial m^2_i\over \partial S}
	\left(\ln m^2_i - 1\right)\over \sum_i g_i m^2_i {\partial m^2_i\over
	\partial S}} \\
\label{bigA}
	A &=& -{1\over 32\pi^2\langle H\rangle} \sum_i g_i m^2_i
	 {\partial m^2_i\over \partial H}
	\left(\ln {m^2_i\over\mu^2} - 1\right)
\eeqa
evaluated at the minimum of the tree level potential.  These
conditions ensure that the position of this minimum remains 
unchanged at one loop.  

\section{Beta functions}
\label{betafns}

Defining $\beta_\lambda = 16\pi^2 d\lambda/d\ln\mu^2$, the beta
functions for the largest couplings in the singlet Majoron model
are \cite{gdm}

\begin{eqnarray}
\beta_{\lambda_h} & = & 12 \lambda_h^2 + \frac12 \lambda_{hs}^2
    + \frac94 g^4 + \frac98 (g^2+g'^2)^2 - 3 y_t^4\nonumber\\
    &+&\lambda_h \left( -\frac92 g^2-\frac32 g'^2 + 6y_t^2 \right)                 \\
\beta_{\lambda_{hs}} & = & 6 \lambda_h \lambda_{hs} + 2 \lambda_{hs}^2
    + 4 \lambda_{hs} \lambda_s
    + \lambda_{hs} \left( -\frac94 g^2-\frac34 g'^2 + 3y_t^2 + 
	\frac12\sum y_i^2 \right)   \\
\beta_{\lambda_s} & = & 2 \lambda_{hs} \lambda_s + 10 \lambda_s^2
    - \frac12 \sum y_i^4 + \lambda_s \sum y_i^2                      \\
\beta_{y} & = & \frac34  y_t^3 + \frac12 y_t \left( 3 y_t^2 - \frac{5}{12}g'^2 
- \frac94 g^2  -8 g_s^2 \right)                \\
\beta_{y_j} & = & \frac18 y_j^3 + \frac14 y_j \sum y_i^2 
\end{eqnarray}

We integrate these starting from the scale $\mu = $100 GeV up to
the first Landau pole (where any of the running couplings diverge),
taking 3 generations of right-handed neutrinos.
The running of the gauge couplings is neglected in this estimate.

\section{Electroweak precision observables}
\label{EWPOS}

To evaluate the impact of constraints on the oblique parameters
$S,T,U$, we follow the procedure of 
references \cite{Profumo:2007wc,Barger:2007im}, defining 
$\Delta\chi^2$ as in eq.\ (5.4) of \cite{Profumo:2007wc}, and
taking points with $\Delta\chi^2>7.8$ to be excluded at 95\% c.l.
Explicit expressions for $T$ were given in those references, but
not for $S$ or $U$.  These can be derived from the definitions found
in eqs.\ (10.61) of the PDG Review of Particle Properties
\cite{RPP}, and the expressions for the $W$ and $Z$ self-energies
in appendix A of \cite{Barger:2007im}.  We find that the contribution
to $S$ from the Higgs sector is
\begin{eqnarray}
-2\pi S_{\rm new} &=& {\cos^2\theta}\left( {1\over m^2_Z}\left[G({H'},Z)
	-m^2_{H'}F_1(Z,H',Z)\right] +F_2(H',Z,Z)\right.\nonumber\\ 
	&+& \left.
	2\left[ F_0(W,H',0)-F_0(W,H',W) + F_0(Z,H',Z)- F_0(Z,H',0)
	\right]\phantom{1\over m^2_Z}\!\!\!\!\!\!\!\! \right)
	\nonumber\\
	&+&\sin^2\theta\left(\phantom{1\over m^2_Z}\!\!\!\!\!\!\!\!
	H'\to S'\right) 
\end{eqnarray} 
where 
\begin{eqnarray}
	G(a,b) &=&  \frac14 \left(m^2_a+m^2_b\right)-
{m^2_a m^2_b\over 2(m^2_a-m^2_b)}\ln{m^2_a\over m^2_b}
-	
     \frac12 \left(m^2_a\ln m^2_a + m^2_b\ln m^2_b\right)\\
	F_n(a,b,c) &=& \int_0^1 dx\, x^n \ln((1-x) m^2_a + +x\, m^2_b
	-x(1-x) m^2_c)
\end{eqnarray}
and we define $m^2_0\equiv 0$. The deviation from the standard model
prediction is obtained by taking
\beq
	\Delta S = S_{\rm new} - S_{\rm new}(\phi=0,\,m^2_{H'}=
	m^2_{S'} = m^2_h)
\label{dS}
\eeq

Similarly, $U$ can be inferred from the combination
\begin{eqnarray}
2\pi (S_{\rm new}+U_{\rm new}) &=& {\cos^2\theta}\left( {m^2_{H'}\over m^2_W}
	\left[F_1(W,H',0)-F_1(W,H',Z,W)\right] +F(H',W,0)\right.\nonumber\\ 
	&+& \left.
	2\left[ F_0(W,H',W)-F_0(W,H',0)
	\right] + F_1(H',W,0)-F_1(H',W,W)
\phantom{1\over m^2_Z}\!\!\!\!\!\!\!\! \right)
	\nonumber\\
	&+&\sin^2\theta\left(\phantom{1\over m^2_Z}\!\!\!\!\!\!\!\!
	H'\to S'\right) 
\end{eqnarray} 
where $F=F_1-F_2$.  $\Delta U$ is computed analogously to (\ref{dS}).


\end{document}